\documentclass[bibyear]{aa} 
\usepackage{color}
\usepackage{graphicx}
\usepackage{txfonts}
\usepackage{enumitem}
\usepackage[utf8]{inputenc} 
\usepackage{cancel}
\usepackage{graphicx}
\usepackage{xcolor}
\usepackage{geometry}
\usepackage[fleqn]{amsmath}
\usepackage{mathtools}
\usepackage{epstopdf}
\usepackage{multirow}
\usepackage{txfonts}
\usepackage{expdlist}
\usepackage{natbib}
\usepackage{color}
\usepackage{epsfig}
\usepackage{url}
\usepackage{placeins}
\usepackage{threeparttable} 
\usepackage{comment}
\usepackage{graphicx}
\usepackage{txfonts}
\newcommand{\swiftJ}[1]{Swift J1727.8–1613{}#1}
\newcommand{\swift}[1]{Swift J1727{}#1}

\begin{document}

   \title{Evolution of the Comptonizing medium of the black-hole candidate Swift J1727.8–1613  along the hard to hard-intermediate state transition using NICER}

   \author{Divya Rawat
          \inst{1},
          Mariano M\'endez\inst{2},
          Federico Garc\'ia\inst{3},
          Pierre Maggi\inst{1}
          }

   \institute{Observatoire Astronomique de Strasbourg, Universit\'e de Strasbourg,
              CNRS, 11 rue de l’Universit\'e, F-67000 Strasbourg, France\\
              \email{rawat@unistra.fr}
         \and
             Kapteyn Astronomical Institute, University of Groningen, PO BOX 800, Groningen NL-9700 AV, the Netherlands\\
             \email{mariano@astro.rug.nl}
         \and
             Instituto Argentino de Radioastronom\'ia (CCT La Plata, CONICET; CICPBA; UNLP), C.C.5, (1894) Villa Elisa, Buenos Aires, Argentina\\
              }

\abstract{We analyse the properties of the Comptonizing medium in the black-hole X-ray binary Swift J1727.8$-$1613 using the time-dependent Comptonization model vkompth, using NICER observations of type-C QPOs in the hard and hard-intermediate states. During the 2023 outburst of the source, we measure the rms and phase lags of the QPO across 45 observations as the QPO frequency, $\nu_{\rm QPO}$, evolves from $\sim 0.3$ Hz to $\sim 7$ Hz. By simultaneously fitting the time-averaged spectrum of the source and the rms and lag spectra of the QPO, we derive the evolution of the disk and corona parameters. At $\nu_{\rm QPO} = 0.34$ Hz, the QPO phase lags are hard, with 10 keV photons lagging 0.5 keV photons by $\sim 0.5$ rad. As $\nu_{\rm QPO}$ increases, the lags for the same energy bands decrease, reaching near zero at $\nu_{\rm QPO} \sim 1.2$ Hz, and then reverse to soft lags of $\sim -1.1$ rad at $\nu_{\rm QPO} \sim 7$ Hz. Initially, the inner radius of the accretion disk is truncated at $\sim 30-40 R_g$ (assuming a 10 solar-mass black hole) and, as the QPO frequency increases, the truncation radius decreases down to $\sim 10 R_g$. Initially, two coronas of sizes of $\sim 6.5 \times 10^3$ km and $\sim 2 \times 10^3$ km, extend over the disk and are illuminated by different regions of the disk. As the QPO frequency increases, both the coronas shrink to $\sim 2 \times 10^3$ km at $\nu_{\rm QPO} = 2.5$ Hz. Following a data gap, one corona expands again, peaking at a size of $\sim 2 \times 10^4$ km. We interpret the evolution of the coronal size in the context of accompanying radio observations, discussing its implications for the interplay between the corona and the jet.}    

\keywords{X-rays: binaries --
                Black hole physics --
                Accretion, accretion disks --
                X-rays: individual:Swift J1727.8–1613 
                Stars: black holes
               }
\authorrunning{Rawat et al. 2025}
\titlerunning{Comptonizing medium of \swiftJ{}}
   \maketitle
\section{Introduction}
Transient black hole X-ray binary (BHXB) systems generally remain in a quiescent state, exhibiting low X-ray flux levels. Before entering into outburst, they often show optical variability, which is seen as a precursor to the impending outburst \citep[e.g.,][]{ha97}. During the outburst phase, these systems undergo notable X-ray flux variability across various energy bands \citep[see, e.g.,][]{re06,mo12}. The X-ray spectrum of these sources typically features a disk black-body component that dominates the low energies \citep{sh73}, along with a power-law component that can extend up to $\sim$1 MeV \citep{ca06}. This non-thermal component is believed to result from the inverse Comptonization of seed-photons from the accretion disk \citep[e.g.][]{gi97}. Moreover, the spectra of these sources often include an iron emission line in the 6–7 keV range, the profile of which is used to probe the black-hole spin \citep[e.g.][]{re08}. Additionally, a Compton reflection hump is commonly observed around 20–30 keV \citep{ba74,ma95}. Since these spectral components dominate distinct energy ranges, variability in flux or photon counts across these bands is frequently analysed to interpret changes in the system's accretion state.

Timing studies of various BHXBs have shown that these systems exhibit a hysteresis loop in the Hardness-Intensity Diagram (HID) during outbursts \citep[e.g.][]{ho01}. The accretion states are classified based on the source's position within the HID. In this context, the Low-Hard State (LHS) is identified by a high hardness ratio and low flux, while the High-Soft State (HSS) features high flux and a low hardness ratio \citep[][and references therein]{be05,do07,be11}. Additionally, two intermediate states in between these two extreme states are recognized as the Hard Intermediate State (HIMS) and the Soft Intermediate State (SIMS). In the HSS the X-ray spectrum is dominated by blackbody disk emission, whereas the non-thermal component prevails in the LHS. The LHS spectrum is characterised by a power law with a photon index of approximately 1.5–2.0 \citep{gi10}, while the HSS includes a softer power-law component with a photon index of $\Gamma \ge 2.5$ \citep{me97,re06,do07}.

BHXBs exhibit variability in their X-ray light curves, which we analyse in the Fourier domain using Power Density Spectra \citep[PDS, see][]{va85}. The PDS reveal significant variability in BHXB across a wide range of timescales; the most important variability components are the Quasi-periodic Oscillations \citep{chen_1997,taki_1997,psaltis_1999,nowak_2000}, categorised into mHz QPOs, Low-Frequency QPOs  (LFQPOs), and High-Frequency QPOs \citep[for a review, see][and references therein]{in19}.  LFQPOs occurring between 0.1 and 30 Hz, are further classified as Type-A, Type-B, and Type-C (\citealt{wi99}, \citealt{ho01}, \citealt{re02}, \citealt{ca04}). Type-C QPOs are the most frequently observed, typically appearing in the LHS and HIMS \citep[see][and references therein]{mo15}. In the PDS, these QPOs feature a large fractional rms amplitude, often accompanied by a harmonic component at twice the QPO frequency, along with a broad Lorentzian noise component and occasionally a sub-harmonic component \citep{belloni_2002,ca04}.
Transient BXBs trace an oval-shaped wheel in the Power Colour-Colour diagram, where the angle along the wheel (the hue) can be used to identify the canonical accretion states of these systems \citep[see][for details]{heil_2015}.

Despite their prevalence in BHXB, the origin of these QPOs remains a subject of debate within the X-ray astronomy community, particularly between two prevailing models. One class of model considers that the LFQPOs are due to a geometric mechanism, as described by the relativistic precession model (RPM; \citealt{st98,st99}), where the type-C QPO frequency would correspond to the Lense-Thirring precession frequency around the black hole \citep{le18}. The RPM has been tested in BHXB systems such as GRO J1655–40 \citep{mo14a}, XTE J1550-564 \citep{mo14b}, and MAXI J1820+070 \citep[][and references therein]{bh21}. An alternative class of models suggests a dynamical rather than a geometric mechanism for LFQPOs,
including accretion-ejection instabilities in a magnetized disc \citep{ta99}, oscillations in a transition layer within the accretion flow \citep{ti04}, and corona oscillations driven by magnetoacoustic waves \citep{ca10,on11}. In this context, LFQPOs have been associated with the relativistic dynamic frequency of a truncated accretion disk in BHXB such as GRS 1915+105 \citep{mi20,li21}, MAXI J1535--571 and H1743--322 \citep{ra23b}, GX 339--4, and EXO 1846--031 \citep{zuo24}.

The timing and spectral properties of X-ray binaries have been widely used to investigate the physical and geometric characteristics of disk-corona systems, specifically through phase lags and rms amplitude spectra at QPO frequencies \citep{le98,le01,ku14}. This approach assumes that the flux oscillations in the time-averaged spectrum—modulated at the QPO frequency—are driven by oscillations in the thermodynamic properties of the Comptonizing region. These properties include the temperature and external heating rate of the corona, the electron density within the corona, and the temperature of the soft photon source, which provides seed-photons for Comptonization. \citet{ka20} introduced {\tt vkompth}, an improved model building on the earlier model by \citet{ku14} with enhanced equation-solving techniques. Originally designed for neutron star binaries, this model was later adapted for black-hole binaries as well \citep{be22}. While this model describes the radiative properties of the QPO (rms amplitude and lags) in terms of feedback loop between the corona and the disk, the same feedback mechanism can explain the QPO frequency \citep{mas22}. To date, the model has effectively explained comptonizing medium properties through type-A, type-B, and/or type-C QPOs in BHXB systems, including GRS 1915+105 \citep{me22,ga22}, MAXI J1535--571 \citep{zh22,Rawat_23a,zh23}, GRO J1655-40 \citep{ro23}, MAXI J1820+070 \citep{ma23}, and MAXI J1348$-$630 \citep{ga21,zh_23_type_A,al24}. 

\swiftJ{} (\swift{} hereafter) is a BHXB source first detected on August 24 2023 with Swift/BAT \citep{page_2023} and initially classified as a GRB source but later identified as a BHXB source with MAXI/GSC \citep{negro_2023a} and NICER \citep{connor_2023}. Using the GTC-10.4m telescope, \citet{mata_2024} proposed that the companion star is an early K-type star and estimated an orbital period of $\sim$ 7.6 h and a distance to the source of  2.7 $\pm$ 0.3 kpc. With the Very Long Baseline Array and the Long Baseline Array, \citet{wood_2024} imaged \swift{} during the hard/hard-intermediate state, revealing a bright core and a large, two-sided, asymmetrical, resolved jet. The polarization properties of \swift{} indicate that the source has an inclination angle between $30-60^{\circ}$ \citep{ve23}.\\

During the initial phase of the outburst, observations with INTEGRAL, ART-XC and Swift/XRT revealed a variable QPO in \swift{}, with frequencies ranging from 0.8 Hz to 10.0 Hz, as reported by \citet{Mereminskiy_24}. With IXPE, a QPO with a frequency evolving from approximately 1.3 Hz to 8.0 Hz was also reported in the 2--8 keV energy band by \citet{ingram_2024}. Observations with IXPE and NICER from August to September 2023 showed that the source transitioned from the LHS to the HIMS, accompanied by a significant decrease in the polarization fraction from about 4\% to 3\% \citep{ingram_2024}. Using INTEGRAL/IBIS, \citet{bouchet_24} reported a highly polarized spectral component above 210 keV, present in both the HIMS and the initial phases of the SIMS. Interestingly, during the HIMS, the polarization angle differs from the angle of the compact jet projected onto the sky, while in the SIMS the two angles are closely aligned \citep{bouchet_24}. During the decay of the outburst, back in the LHS (during February 2024), the polarization angle (PA) remained constant at the same value as at the one during the rising part, while the polarization fraction increased with X-ray hardness \citep[4--8 keV/2--4 keV;][]{Podgorny_2024}. 

In this work, we analyse NICER observations of \swift{} during its LHS and HIMS to study the Comptonizing medium properties through the type-C QPOs using the {\tt vkompth} model. In Section \ref{sec:obs} we provide details of the observations, instrument specifications, and the spectral and timing analysis techniques applied. The fits to the rms, lag and time-averaged spectra at the type-C QPO are presented in Section \ref{sec:Res}. We discuss the implications of these findings and interpret them within the disk-corona framework in Section \ref{sec:dis}.

\begin{figure}
\hspace*{-.2cm}\includegraphics[scale=0.46]{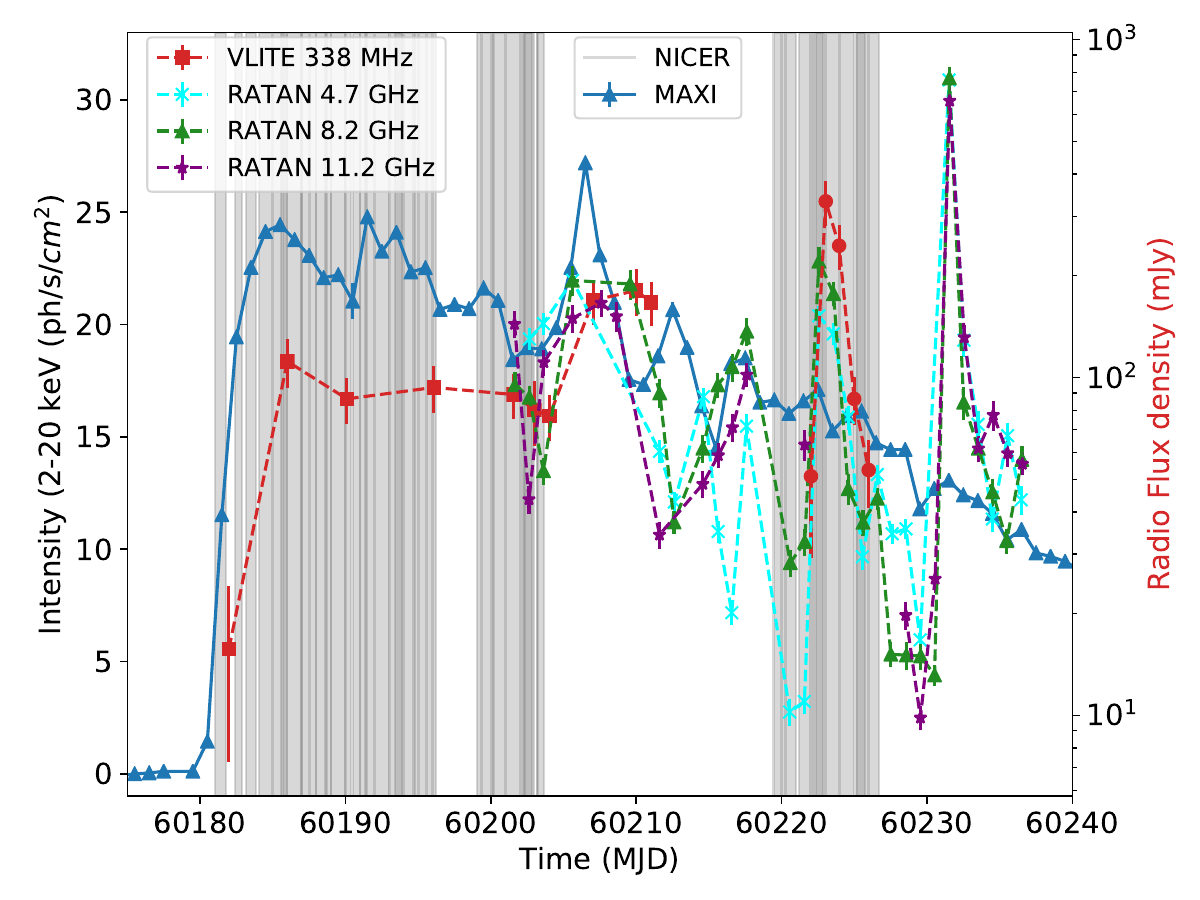}
\caption{The MAXI light curve of \swiftJ{} starting from 60175 MJD (August 19, 2023) in the 2--20 keV energy band in units of photons s$^{-1}$ cm$^{-2}$ (blue triangles; left $y$ axis).  The right $y$-axis shows the radio flux density in mJy from the VLA Low-band Ionosphere and Transient Experiment (VLITE) at 338 MHz \citep[red data points, data from][]{peter_2023ATel}, alongside RATAN-600 observations at frequencies of 4.7 GHz (cyan data points), 8.2 GHz (green data points), and 11.2 GHz \citep[purple data points, data from][]{ingram_2024}.}
\label{lightcurve}
\end{figure}

\begin{figure}
\centering\includegraphics[scale=0.485]{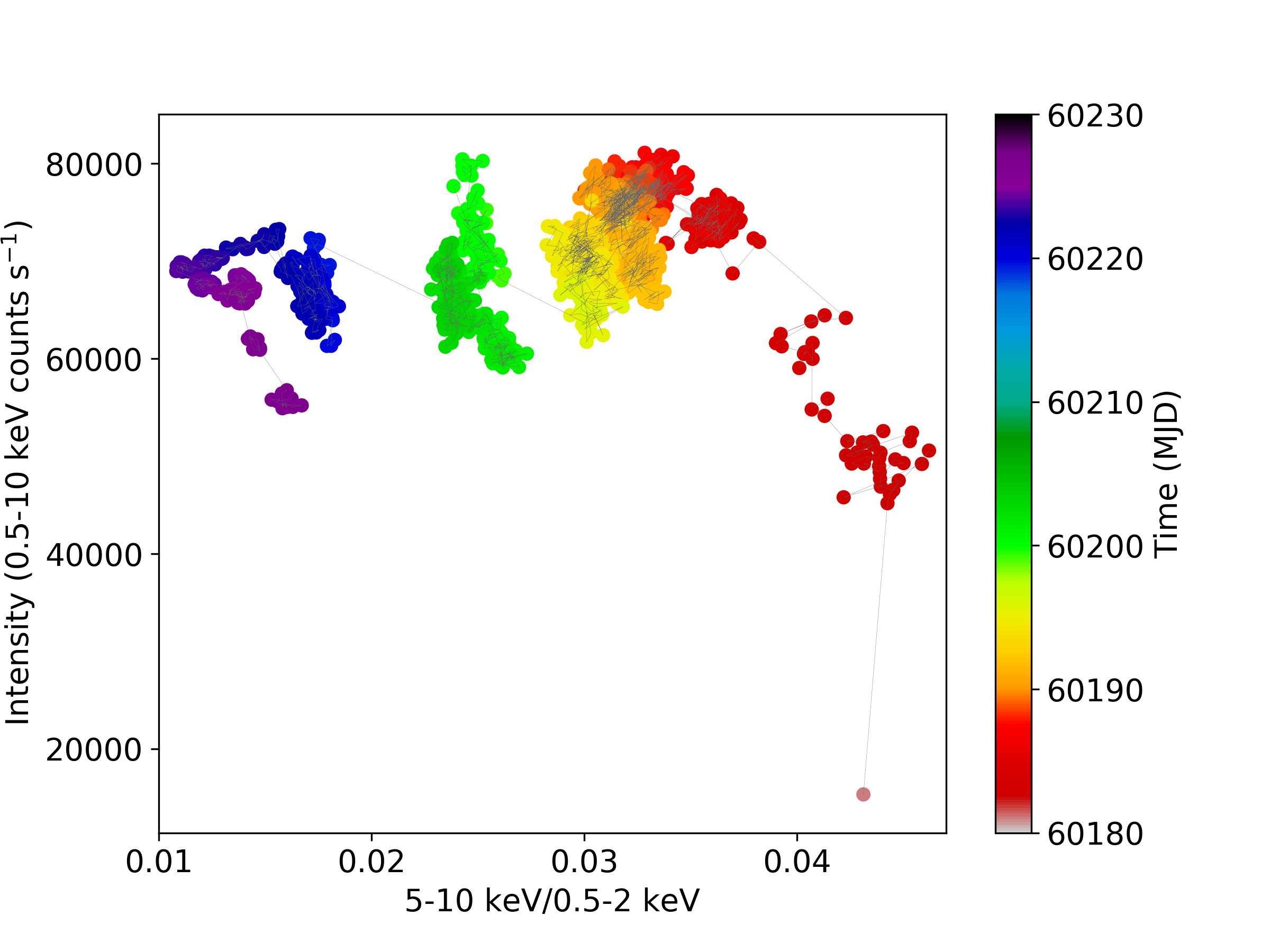}
\caption{The NICER hardness-intensity diagram starting from  MJD 60181. The intensity is the 0.5--10.0 keV count rate and the hardness is the ratio of the count rate in the 5--10 keV band to that in the 0.5--2.0 keV band, with data points binned over 100 seconds. The colour scale indicates the MJD.}
\label{HID}
\end{figure}

\begin{figure*}
    \rotatebox{-90}{\includegraphics[scale=0.65]{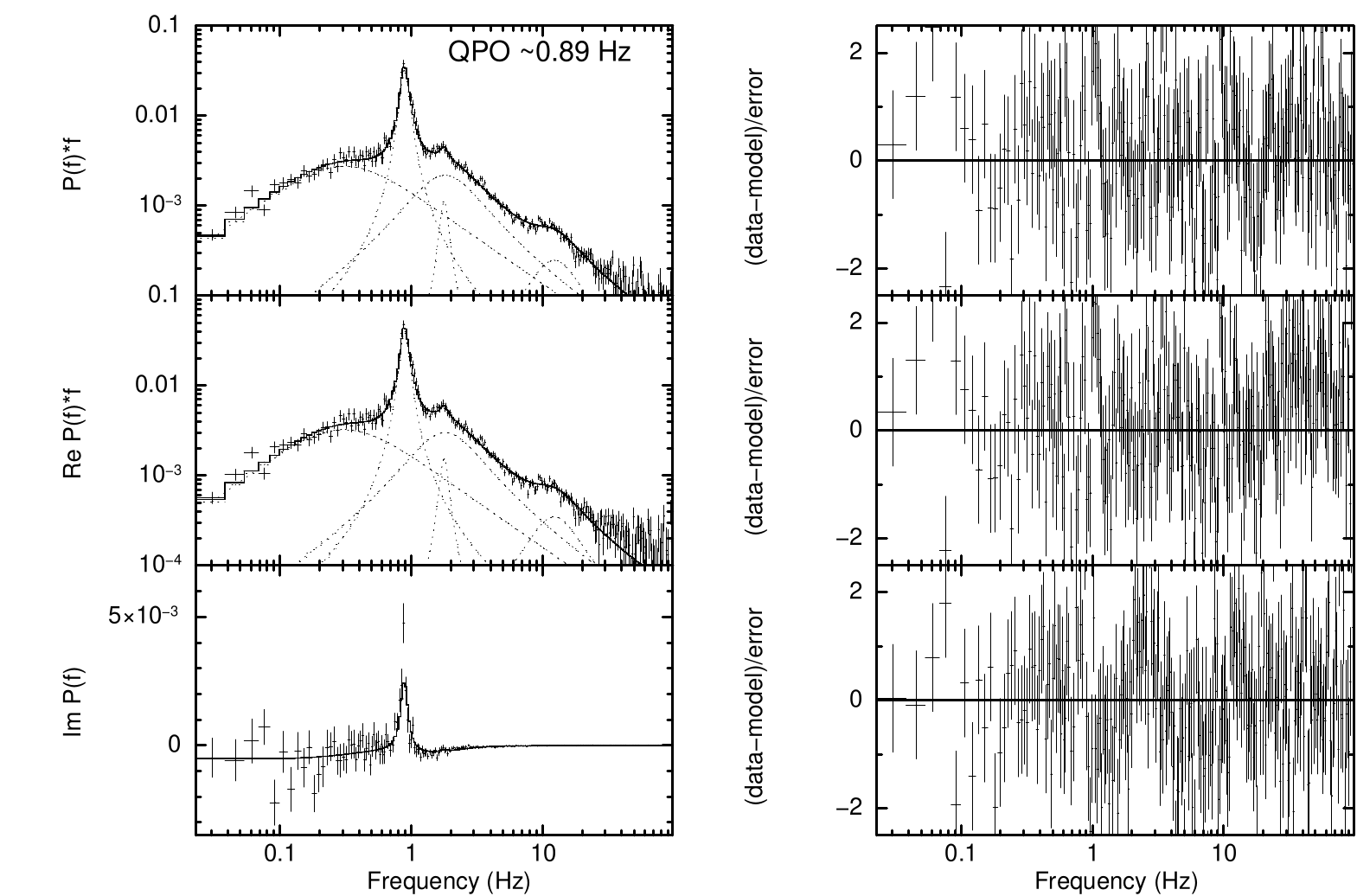}}

\caption{The top left panel displays the 0.5–10.0 keV power density spectrum of \swiftJ{}, fitted with five Lorentzians for ObsID 6203980105. The middle and bottom left panels show the real and imaginary parts of the cross spectrum between the 0.5--2.0 keV and 2.0--10.0 keV bands, along with the best-fitting model that assumes that the phase lags of each Lorentzians are constant with Fourier frequency \citep[see][]{me_2024}.  The right-hand panels display the residuals of the fits for the respective spectra.}

\label{PDS_real_imag}
\end{figure*}

\section{Observations and data analysis}
\label{sec:obs}
The Neutron Star Interior Composition Explorer (NICER), launched on June 2, 2017, is mounted on the International Space Station \citep[ISS;][]{ge16}. Its X-ray Timing Instrument (XTI) operates within the 0.2–12.0 keV energy range, with an effective area exceeding 2000 cm$^{2}$ at 1.5 keV. NICER provides an energy resolution of 85 eV at 1 keV and a time resolution of 4 × 10$^{-8}$ s. We used NICER observations of \swift{} from 2023-08-26 to 2023-10-09, with observation details presented in Table \ref{table1}.\

 We processed each observation using the {\tt{nicerl2}} task, applying standard calibration and screening procedures. By default, {\tt{nicerl2}} uses the `night' threshold setting, but there are observations with no exposure during the night; for those observations, we set {\tt{threshfilter=DAY}}. We used either day or night data for an individual observation, because NICER has been affected by optical light leakage on May 2023 and hence separate analysis of orbit day and orbit night data is recommended by the NICER team\footnote{\url{https://heasarc.gsfc.nasa.gov/docs/nicer/analysis_threads/nicerl2/}}. We only used observations where the exposure time is greater than 200 seconds after running the  {\tt{nicerl2}} task. Next, we extracted light curves in the 0.3–4.0 keV, 8.0–12.0 keV, and 0.5–10.0 keV energy bands using the {\tt{nicerl3-lc}} task. The left panel of Figure \ref{lightcurve} shows the MAXI light curve for \swift{}, with NICER observations highlighted in the grey shaded regions. Details of the NICER count rates are provided in Appendix Table \ref{table1}. We then calculated the hardness ratio as the number of photons in the 5.0–10.0 keV band divided by those in the 0.5–2.0 keV band. We segment the observations where a significant change in both the full-band count rate and hardness ratio was observed. To illustrate the source evolution, we plot the HID using NICER data in the Figure \ref{HID}. The colour scale on the right side of that Figure represents time, with red corresponding to MJD 60181 and transitioning to navy blue at MJD 60226, 45 days later.
 
\subsection{Timing analysis}
We extracted the fractional rms  normalised \citep{be90} PDS and cross spectrum for each observation using the General High-energy Aperiodic Timing Software (GHATS)\footnote{\url{http://www.brera.inaf.it/utenti/belloni/GHATS_Package/Home.html}} V3.1.0. We used the novel technique proposed by \citet{me_2024} to measure the phase lags.  For each observation, we first extracted the 0.5--10.0 keV PDS and the cross spectrum between the 0.5--2.0 keV and 2.0--10.0 keV energy bands. We fit the PDS with a combination of Lorentzian functions together with the real and imaginary parts of the cross spectrum. For each component present in the PDS, we fit the real and imaginary part of the cross spectrum with Lorentzians multiplied by cos($\Delta \phi$) and sin($\Delta \phi$), respectively with the centroid and width of each Lorentzian tied to the parameters in the PDS, where $\Delta \phi$ is the phase lag. 

We observe that the QPO centroid frequency varies from  0.3 Hz to 7.0 Hz (Appendix Table \ref{table1}). For some observations marked with a $^*$ in Appendix Table \ref{table1}, the QPO detection was below 3 sigma, so these observations were excluded from further analysis. Additionally, for obs no. 3 and 4, the QPO was weak in the 0.5–10.0 keV band, and therefore, these observations were also excluded from further study. To measure the phase-lag and rms spectra, we divide the energy bands into 38 bins, starting from 0.5 keV to 7.9 keV with a bin size of 0.2 keV, and a last bin that spans from 7.9 keV to 10 keV due to the QPO being weaker in the highest energy band. For some observations (mainly those where we segmented the observation into parts or the QPO frequency exceeded 2.5 Hz), the QPO fractional rms was low (as shown in Appendix Table \ref{table1}), making the QPO insignificant with finer binning. To address this in those observations, we manually merge the bins in regions where the signal was weak, resulting in 11 energy bands: 0.5--0.9 keV, 0.9--1.3 keV, 1.3--1.7 keV, 1.7--2.1 keV, 2.1--2.9 keV, 2.9--3.7 keV, 3.7--4.5 keV, 4.5--5.7 keV, 5.7--6.5 keV, 6.5--7.7 keV, and 7.7--10.0 keV. For each band we extract the phase lag and rms using the technique of \citet{me_2024} discussed above using the 0.5--10.0 keV band as the reference. To account for the correlation due to photons being present in both the subject and reference bands, we include a constant term in the model of the real part of the cross-spectrum.

 \subsection{Spectral analysis}
 \label{section:spectral_analysis}
 We use the nicerl3-spect task to extract source and background spectra, response (rmf) and ancillary response (arf) files in the 0.5--10.0 keV band. We used {\tt{Heasoft}} version 6.33 and CALDB version 20240206.  We fitted the  background-subtracted time-averaged spectrum of the source using the model  {\tt TBfeo*(diskbb + gaussian + nthComp)} in {\tt{Xspec 12.14.0}}. The {\tt{TBfeo}} model is similar to the {\tt{tbabs}} model, which accounts for interstellar absorption, but allows the oxygen and iron abundances to vary, in addition to the hydrogen column density. We used the cross-section tables of \citet{ve96} and the abundances of \citet{wi00} and left the hydrogen column density as a free parameter. We have frozen the oxygen and iron abundance relative to Solar to 1 and the redshift to zero. The {\tt{diskbb}} component models the thermal emission from an optically thick and geometrically thin accretion disk \citep{mi84,ma86} while {\tt{nthcomp}} \citep{zd96,zy99} models the Comptonised emission from the X-ray corona. We kept both {\tt{diskbb}} parameters, the temperature at the inner disk radius, $kT_{in}$, and the normalisation free. The  {\tt{nthcomp}} model parameters are the power-law photon index, $\Gamma$, electron temperature, $kT_e$, seed-photon temperature,  $kT_{bb}$, and normalization. The seed-photon temperature $kT_{bb}$ was tied to $kT_{in}$ of the {\tt{diskbb}} component. Since the electron temperature of {\tt{nthcomp}} could not be constrained with the NICER data, following \citet{bouchet_24} we fixed it at 40 keV for observations 2 to 36 and at 250 keV for observations 37 to 54. We modelled the relatively broad iron line present in the residuals of observations 1 to 17 with a Gaussian fixed at 6.4 keV using the {\tt {gauss}} model. 

\begin{figure}
\hspace*{-.2cm}\includegraphics[scale=0.48]{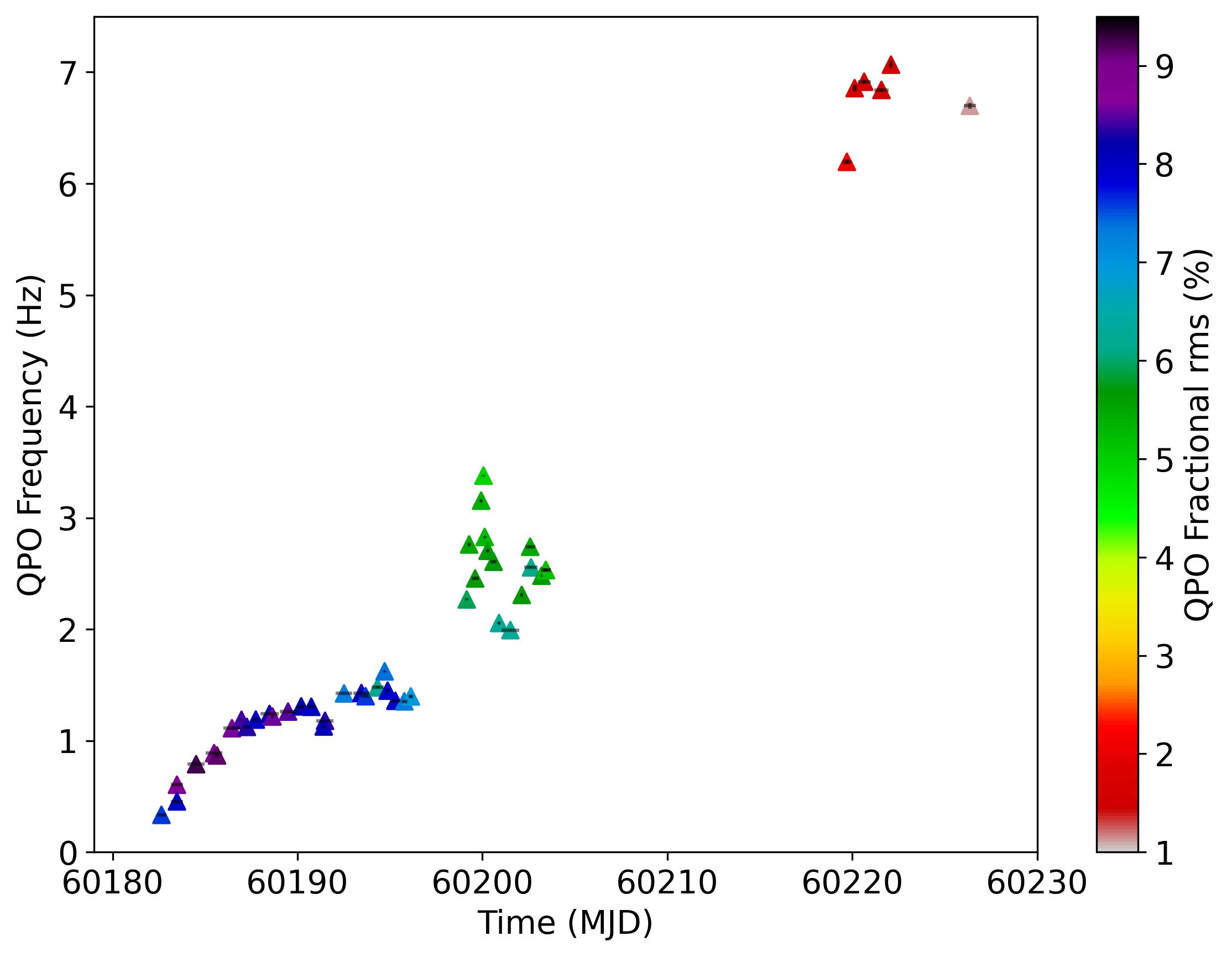}
\caption{QPO frequency of \swiftJ{} as a function of MJD. The colour bar gives the QPO fractional rms amplitude in the 0.5--10.0 keV energy range.}
\label{qpo_fract}
\end{figure}


\begin{figure}
\hspace*{-.2cm}
\includegraphics[scale=0.62]{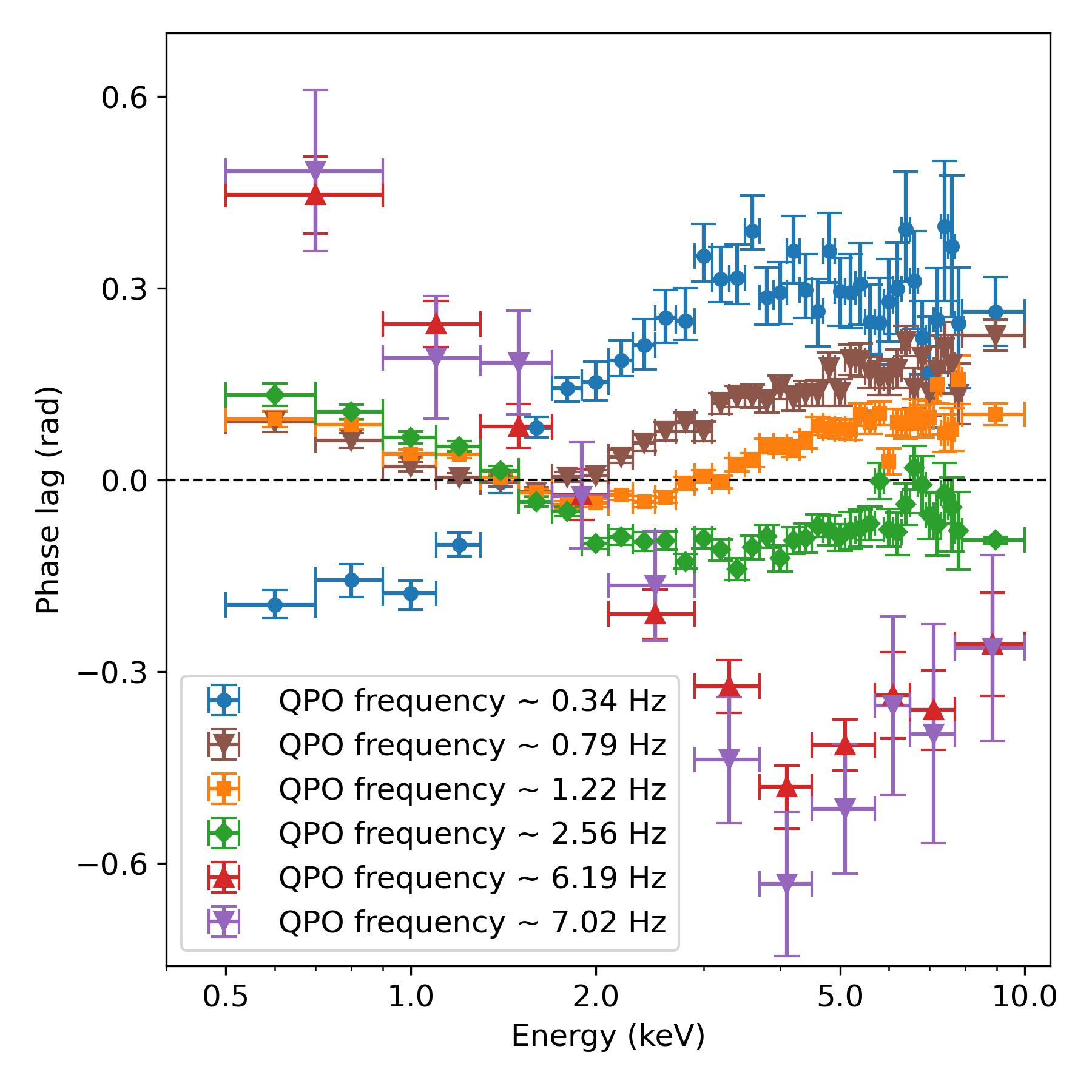}
\caption{The phase-lag spectra corresponding to specific QPO frequencies, from 0.34 Hz to 7.02 Hz. The reference energy band used here for the lags is 0.5--10.0 keV.}
\label{phase_lag_all}
\end{figure}
\begin{table}
    \begin{center}
    \renewcommand\arraystretch{1.4}
	\caption{Representative best-fitting spectral and corona parameters for ObsID 6203980105 of \swiftJ{} using the single and dual corona model.} 
	\label{table2}
	\resizebox{\columnwidth}{!}{
	\begin{tabular}{lccc} 
		\hline
		Component & Parameter & M1$^{\rm (1)}$ & M2$^{\rm (2)}$\\
		\hline
		TBfeo& $N_{\rm H} (10^{22}\,\rm{cm^{-2}})$ & [0.28]$^{a}$ & [0.28]$^{a}$ \\
		diskbb & $kT_{\rm in}$ (keV) & $0.351\pm{0.003}$ & $0.352\pm{0.003}$\\
	    & $N_{\rm disk}\,(10^{4})$   & $15.8 \pm {0.2}$ & $15.7\pm{0.4}$\\
		nthComp & $\Gamma$ & $1.742\pm{0.002}$ & $1.742\pm{0.003}$\\
		 & $kT_{\rm e}$ (keV) & [40]$^{a}$ & [40]$^{a}$ \\	
		 & $N_{\rm nthComp}$ & $34.4^{+0.3}_{-0.2}$ & $34.4\pm{0.3}$\\
		gaussian & LineE (keV) & [6.4]$^{a}$ & [6.4]$^{a}$ \\
		 & $\sigma$  (keV) &  $[1.0]^a$ & $[1.0]^a$ \\
		 & Strength ${\rm (10^{-1})}$& $[0.2]^a$ & $[0.2]^a$\\
		vkompthdk/vkdualdk & $kT_{\rm s}$ / $kT_{\rm s,1}$ (keV) & $0.057 \pm 0.001$  & $0.286^{+0.020}_{-0.055}$ \\
		& $L$ / $L_{1}\,{\rm (10^{3}\,km)}$  & $31.4 \pm {0.7}$ & $6.1^{+1.5}_{-1.3}$ \\
		& $L_{2}\,{\rm (10^{3}\,km)}$ & $-$ & $7.5^{+2.1}_{-1.7}$\\
		& $\eta$ / $\eta_{1}$ &  $0.60^{+0.03}_{-0.02}$ & $0.28^{+0.02}_{-0.04}$  \\
		& $\eta_{2}$ & $-$ & $0.32 \pm 0.01$ \\
		& $\phi$ (rad) & $-$  & $3.11\pm{0.01}$ \\
		& $\rm {reflag}$ $(10^{-2} )$ & $3.1\pm{0.2}$  & $8.2 \pm 0.5$ \\	
		&  $\eta_{\rm{int}}$/$\eta_{\rm{int,1}}$ & $0.013 \pm 0.001$ & $0.014^{+0.003}_{-0.002}$   \\
            
		& $\eta_{\rm{int,2}}$ & $-$ & $0.017 \pm 0.001$ \\
		\hline
		& $\chi^{2}$ (dof) & $415.2 \;( 255 \;)$  & $254.2 \;(251 \;)$\\
		
		\hline
	\end{tabular}
	}
	\end{center}
\tablefoot{Uncertainties are given at the $1\sigma$ level.
\tablefoottext{1}{In Model M1, the rms spectra are fitted with {\tt vkompthdk*dilution}, while the lag spectra are fitted with {\tt vkompthdk}.}
\tablefoottext{2}{In Model M2, the rms spectra are fitted with {\tt vkdualdk*dilution}, while the lag spectra are fitted with {\tt vkdualdk}.}
\tablefoottext{a}{The parameters $N_{\rm H}$ and LineE are frozen.}
}	
\end{table}


\begin{figure}
    \centering
    \includegraphics[scale=0.45]{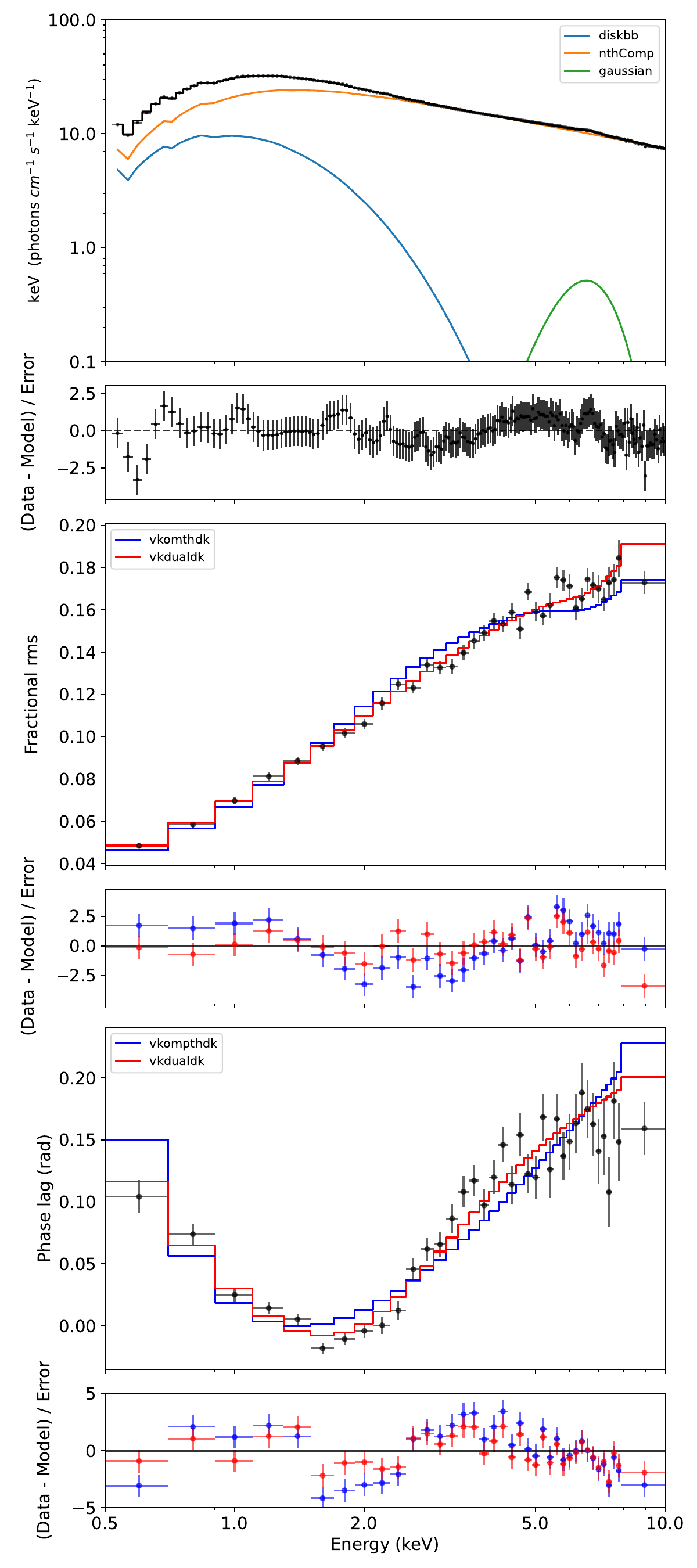}
 
    \caption{The top panel shows the NICER time-averaged spectrum of \swiftJ{} (ObsID 6203980105) with a QPO centroid frequency of 0.89 Hz, fitted using the model {\tt TBfeo*(diskbb + gaussian + nthComp)} together with the residuals of the best-fitting model. The middle panel shows the rms spectrum of the QPO fitted with the {\tt vkdualdk*dilution} (red) and {\tt vkompthdk*dilution} (blue) models with the residuals of the best-fitting model. The bottom panel shows the phase-lag spectrum of at the QPO fitted with the {\tt vkdualdk} (red) and {\tt vkompthdk} model (blue) with the residuals of the best-fitting model. The 0.5–10.0 keV energy band is used as the reference for the phase-lag spectra.}
    \label{spectra_single_dual_corona}
\end{figure}

\subsection{Time-dependent Comptonization model}
To study the properties of the Comptonizing medium, we model the rms and phase-lag spectra of the QPO with the time-dependent Comptonization model {\tt vkompth}\footnote{\url{https://github.com/candebellavita/vkompth}}. The time-averaged variant of {\tt vkompth} model is identical to the thermal Comptonization model {\tt nthComp}, but {\tt vkompth} incorporates additional parameters that only impact the time-dependent version of the model. In the case of a disk-blackbody soft photon-source, the {\tt vkompth} model has two versions: one for a single corona, {\tt vkompthdk}, and second for a dual corona, {\tt vkdualdk}. The model parameters for {\tt vkompthdk}  are the seed-photon temperature, $kT_{\rm s}$, the electron temperature, $kT_{\rm e}$,  the photon index, $\Gamma$, the corona size, $L$, the feedback fraction, $\eta$, the size of the seed-photon source, $a_{f}$, the amplitude of the variability of the external heating rate, $\delta \Dot{H}_{\rm ext}$, and the reference lag of the model in the 0.5--10.0 keV energy band, \textit{reflag}. Since the {\tt vkompth} model is not sensitive to $a_{f}$, we fix this parameter at 250 km \citep[for details, see][]{ga22}.
The parameter {\it{reflag}} is an additive normalisation that allows the model to match the data, given that the observer is free to choose the reference energy band of the lags.
The {\tt vkdualdk}  model \citep{ga21, be22} has two sets of coronal parameters, $kT_{\rm s,1}/kT_{\rm s,2}$, $kT_{\rm e,1}/kT_{\rm e,2}$, $\Gamma_{1}/\Gamma_{2}$, $L_{1}/L_{2}$, $\eta_{1}/\eta_{2}$, and $\delta \Dot{H}_{\rm ext,1}/\delta \Dot{H}_{\rm ext,2}$ to describe the physical properties of the two coronal regions. The remaining parameters of the \textit{dual} models are the size of the seed-photon source and the reference lag, plus an additional parameter, $\phi$, that describes the phase difference between the two coupled coronae \citep[for more details see][]{ga21}. To model the rms spectra of the QPO,we incorporate a multiplicative {\tt dilution} component, which is a function of energy ($E$). This component accounts for the fact that the rms amplitude we observe is diluted by the emission of the other components that we assume do not vary \citep[see details in][]{be22}. We linked the parameters of the {\tt dilution} component to the corresponding parameter of the model of the time-averaged spectrum. We will discuss the results of the fit in detail in Section \ref{subsec:res_spectra}.

\section{Results}
\label{sec:Res}
We present the MAXI lightcurve of \swift{} during its outburst in August 2023 in Figure \ref{lightcurve}. To show the simultaneous evolution of the source in radio, we over-plot the radio light curves obtained from the VLA Low-band Ionosphere and Transient Experiment VLITE 338 MHz \citep[data from][]{peter_2023ATel}, along with the RATAN-600 observations at 4.7 GHz, 8.2 GHz and 11.2 GHz \citep[data from][]{ingram_2024}. The MAXI flux in the 2--20 keV band rises from $\sim$0.5 photons s$^{-1}$ cm$^{-2}$ to $\sim$25 photons s$^{-1}$ cm$^{-2}$ over a period of 5 days, starting from  MJD 60180 (August 24, 2023). The VLITE light curve shows a significant increase in radio flux starting from MJD 60181.98 that remains more or less constant until MJD 60204.04. The X-ray flux then decreases to $\sim$5 photons s$^{-1}$ cm$^{-2}$ over nearly two months, with several X-ray and radio flares occurring in between. The VLITE data show two radio flares on  MJD 60209.99 and MJD 60223.04, while a very bright radio flare is visible in the RATAN-600 data at 4.7 GHz, 8.2 GHz, 11.2 GHz frequencies on MJD 60231.54. We mark the times of the NICER observations with grey regions in Figure \ref{lightcurve}. The NICER $0.5-10.0$ keV count rate initially increases from $\sim$1.5 $\times10^{4}$ counts s$^{-1}$ to $\sim$7.5 $\times 10^{4}$ counts s$^{-1}$. Simultaneously, the HR of the source decreases from $\sim$0.045 to $\sim$0.035, indicating a clear evolution from the LHS to the HIMS as shown in Figure \ref{HID}.
The HR decreases further from approximately $\sim$0.035 to $\sim$0.015, while the flux continues to decrease monotonically, reaching around $\sim$5.7 $\times10^{4}$ counts s$^{-1}$.

\subsection{Timing results}
Adopting the method of \citet{belloni_2002}, we fit the PDS with four to five Lorentzian components. Following \citet{me_2024}, we use the same combination of Lorentzians to fit the real and imaginary parts of the cross spectra multiplied by, respectively, the cosine and sine of the phase lags which, for each individual Lorentzian, are assumed to be constant with frequency. During the fits we link the frequencies and FWHM of each Lorentzian in the PDS and cross spectra. The model includes two components that represent the broadband noise: (1) a zero-centred Lorentzian representing the low frequency broad-band noise and (2) a high-frequency broad noise component around 11.0 Hz. To fit the narrow peak components, which we identify as the QPO and its second harmonic, we use two separate Lorentzians. While fitting the harmonic component, we set the Lorentzian centroid frequency and width to twice the corresponding values of the QPO fundamental. For QPO centroid frequencies less than 3.0 Hz, the PDS also shows a low-frequency noise component (see, for example, the top left panel of Figure \ref{PDS_real_imag}). We include an additional Lorentzian to account for this feature whenever required.\\

We show the fit to the PDS and the real and imaginary part of the cross spectrum of OBSID 6203980105 in Figure \ref{PDS_real_imag}. The top left panel shows the PDS and the right panel shows the residuals of the fit. The PDS shows a significant narrow QPO component at $\sim$0.89 Hz with a harmonic component at twice that frequency. On the basis of strength of the QPO and the presence of a harmonic, and broad noise components, we identify this as a type-C QPO. In the middle left and bottom left panels of Figure \ref{PDS_real_imag}, we display the real and imaginary parts of the cross spectrum, along with the corresponding residuals of the fit in the middle and bottom-right panels. For ObsID 6557020402, which features a QPO at approximately 7.0 Hz, we present the PDS and the real and imaginary parts of the cross spectrum in Appendix Figure \ref{app_PDS_real_imag}. The real part of the cross spectrum contains the same number of Lorentzians as the PDS. Meanwhile, the imaginary part shifts from positive to negative as the QPO frequency increases from 0.3 to 7.0 Hz (see bottom left panel of Figure \ref{PDS_real_imag} and Appendix Figure \ref{app_PDS_real_imag}). \\

The QPO frequency increases monotonically from 0.3 Hz to 1.4 Hz over time until MJD 60196, as shown in Figure \ref{qpo_fract} (see also Appendix Table \ref{table1}). Between MJD 60199 and MJD 60201, the QPO frequency abruptly increases from 2.3 Hz to 3.3 Hz, then drops to 2.0 Hz within a span of just one day. The QPO frequency then increases further to 7.0 Hz, with a gap in NICER data between MJD 60203 and MJD 60219. The colour scale on the right side of the Figure indicates the QPO fractional rms in the $0.5-10.0$ keV band. The QPO fractional rms first increases from 7.6\% to 9.3\% as the QPO frequency increases from $\sim$0.3 Hz to $\sim$0.8 Hz, and then decreases from 9.3\% to $\sim$ 1\% as the QPO frequency increases from $\sim$1 Hz to $\sim$7 Hz at the end of the NICER observations (see also Appendix Table \ref{table1}). During the period, where the QPO frequency exhibits an abrupt increase and then decreases, the fractional rms remains anti correlated. This behaviour of the fractional rms is similar to that observed in other BHXB sources like GRS 1915+105 \citep{zh20}, and MAXI J1535$-$571 \citep{Rawat_23a}. We give details of the QPO centroid frequency, width and fractional rms in the 0.5--10.0 keV band in Appendix Table \ref{table1}.\\
We show the QPO phase lag spectra at five QPO frequencies, 0.34 Hz, 0.79 Hz, 1.22 Hz, 2.56 Hz, 6.19 Hz, and 7.02 Hz, in Figure \ref{phase_lag_all}. At a QPO frequency of 0.34 Hz, the phase lags are negative  at low energies and become positive at $\sim$1.5 keV, continue increasing and become more or less constant at 0.25 rad above $\sim$ 3 keV. The lags increase with energy when $\nu_{QPO} \leq$ 0.34 Hz, flatten for 0.79 Hz $< \nu_{QPO} \leq$ 2.56 Hz, and decrease with energy when $\nu_{QPO} \geq$ 6.19 Hz. Additionally, the lags show minima that shift to higher energies as a function of QPO frequency. At $\nu_{QPO} =$ 0.34 Hz, the minimum appears between 0.5 and 1.0 keV; for 0.79 Hz $< \nu_{QPO} \leq$ 2.56 Hz, the minimum shifts to 1.5--2.0 keV, and by $\nu_{QPO} \geq$ 6.19 Hz, the minimum is at around 4 keV. A similar shift in the phase-lag spectra minima is also observed for the type-B QPOs in MAXI J1348--630 and for the type-C QPOs in MAXI J1535--571 \citep{Rawat_23a}. We also computed the time lag of the 2.0–10.0 keV band with respect to 0.5–2.0 keV band, across all observations (see the second row of Appendix Tables \ref{table3}, \ref{table4}, \ref{table5} and \ref{table6}). 

\begin{figure*}
\centering\includegraphics[scale=0.46]{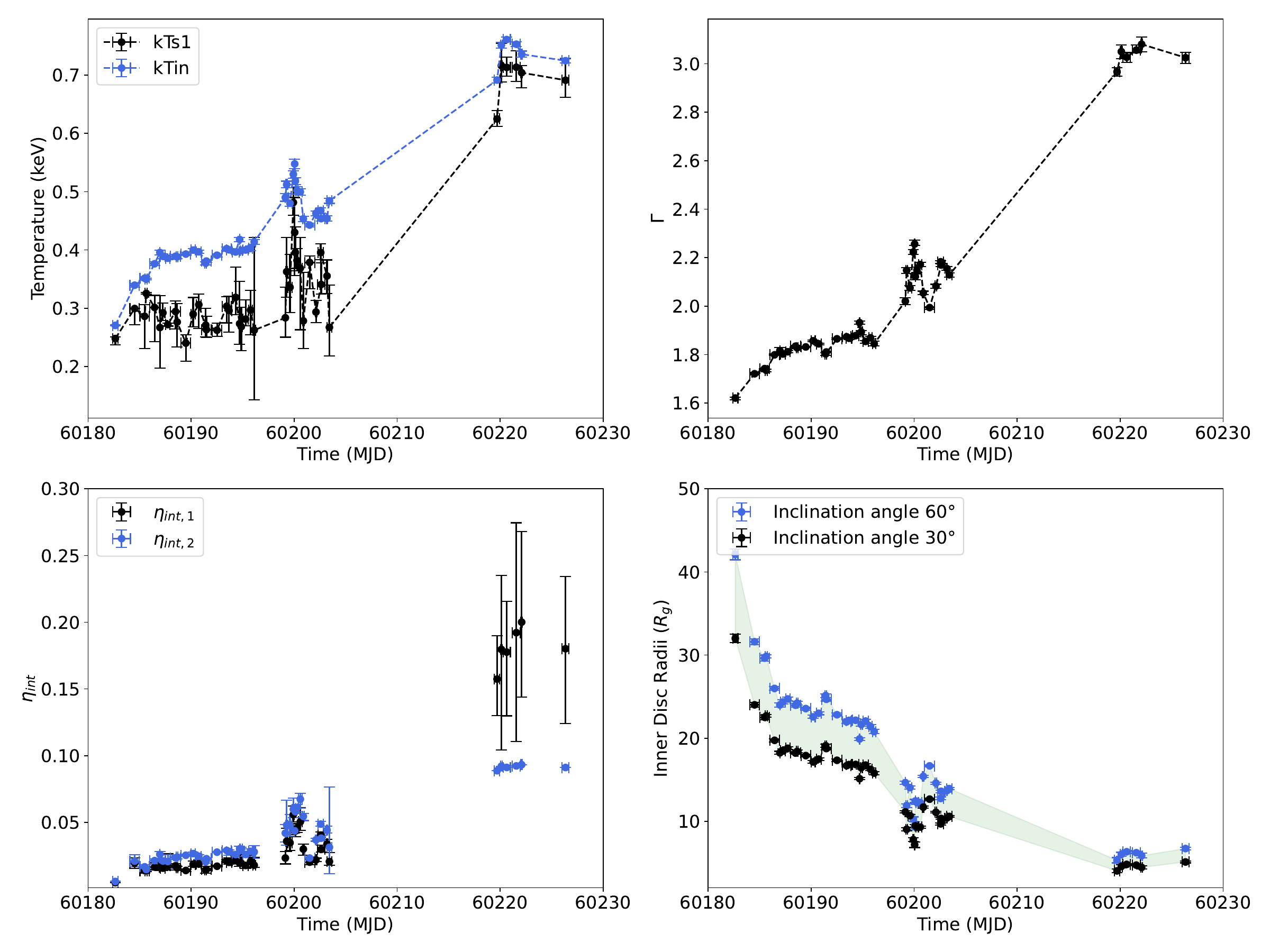}
\caption{The top left panel shows the inner disk temperature, $kT_{in}$, and coronal seed temperature, $kT_{\rm s,1}$, as a function of time for \swiftJ{}. The top right panel presents the power-law photon index of {\sc nthcomp}, $\Gamma$, as a function of time. The bottom left panel displays the time evolution of the intrinsic feedback fraction, $\eta_{\rm{int,1}}$ and  $\eta_{\rm{int,2}}$, obtained from the fit  with the {\tt vkdualdk} model. The bottom right panel shows the evolution of the inner disk radius (in units of gravitational radii) for inclination angles of 60$^\circ$ (blue) and 30$^\circ$ (black). For the calculation of the inner disk radius, we used a spectral hardening factor $T_{col}/T_{eff}$ of 1.7, assumed a M$_\odot$ black hole, and set the distance to the source at 2.7 kpc.}
\label{qpo_para_radio}
\end{figure*}
\begin{figure}
\hspace*{-.2cm}\includegraphics[scale=.46]{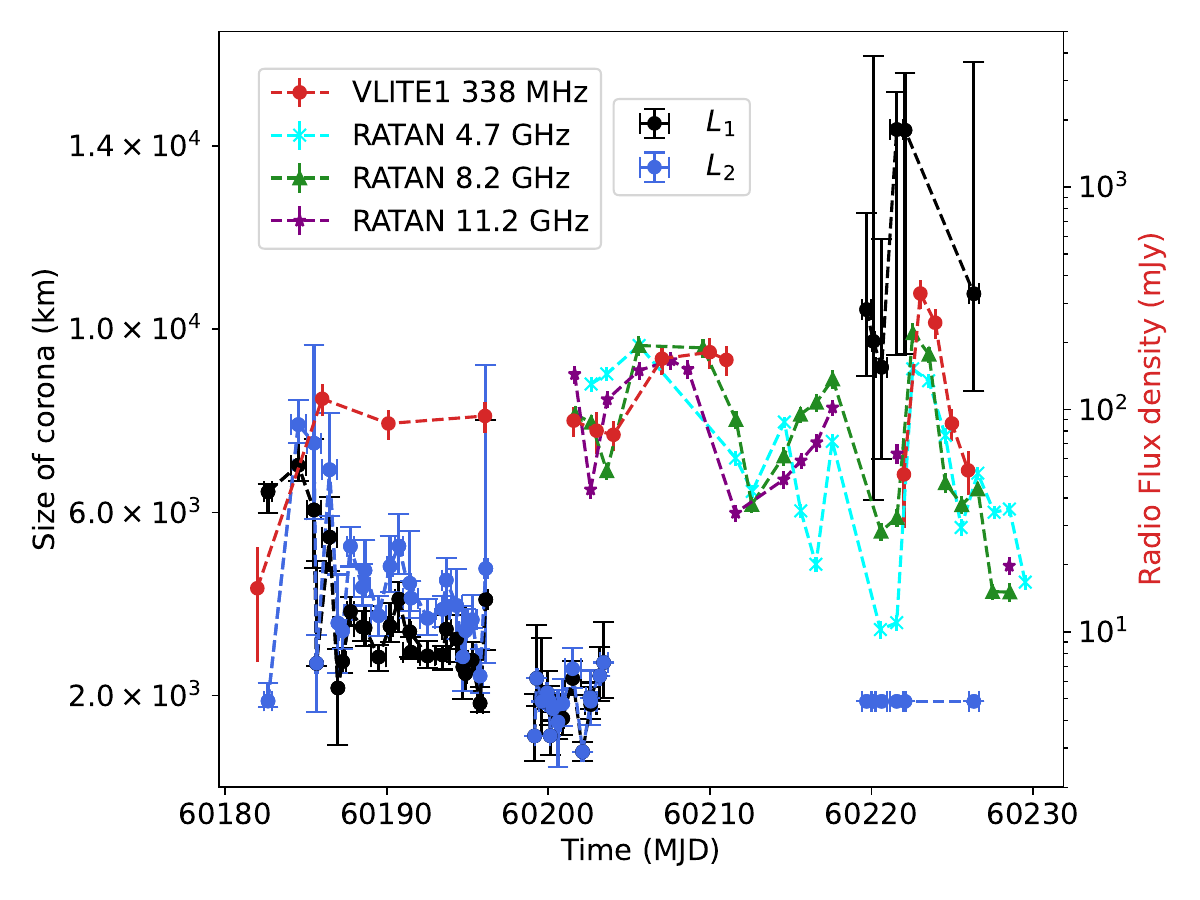}
\caption{The evolution of the sizes of the coronas, $L_1$ and $L_2$, for \swiftJ{} using the {\tt vkdualdk} model. The right $y$-axis shows the radio flux density in mJy obtained from the VLA Low-band Ionosphere and Transient Experiment (VLITE) at 338 MHz \citep[red data points, data from][]{peter_2023ATel}, alongside RATAN-600 observations at frequencies of 4.7 GHz (cyan data points), 8.2 GHz (green data points), and 11.2 GHz \citep[purple data points, data from][]{ingram_2024}.} 
\label{corona_size_radio}
\end{figure}
 
\subsection{Spectral-timing results}\label{subsec:res_spectra}
We fit simultaneously the rms spectrum of the QPO with the model {\tt vkompthdk*dilution}, the phase-lag spectrum of the QPO with the model {\tt vkompthdk} and the time averaged spectrum of the source with model described in Section \ref{section:spectral_analysis}. We tie $kT_{\rm e}$ and $\Gamma$ of {\tt vkompthdk} to  $kT_{\rm e}$ and $\Gamma$ of {\tt nthcomp},  and link all other parameters of {\tt vkompthdk} between the lag spectra and the rms spectra. While fitting the {\tt vkompthdk} model, we initially tied the diskbb temperature ($kT_{in}$) to the seed photon temperature ($kT_s$) of {\tt vkompthdk}. However, the model was unable to reproduce the valley in the phase lag spectrum and gives a $\chi^2$/dof of 1748.1/253. A similar behaviour was observed in the black hole X-ray binary MAXI~J1535--571 when modelled with {\tt vkompthdk}, where the fit improves significantly when $kT_{in}$ and $kT_s$ were untied \citep[see Figure 6 of][]{Rawat_23a}. To address this, we allowed $kT_{in}$ and $kT_s$ to vary independently, and found two separate sources of seed photon; one with a temperature of $kT_{in} \sim 0.35$ keV, responsible for producing the steady-state spectra, and  another with a temperature of $kT_s \sim 0.06$ keV, which contributes to the phase lag and rms spectra. As an example, we show the time averaged spectrum of the source fitted with the model {\tt TBfeo*(diskbb + gaussian + nthComp)}, and the rms and  phase-lag spectra of the QPO at 0.89 Hz fitted with the model {\tt vkompthdk*dilution} and {\tt vkompthdk} in, respectively, the top, middle and bottom panels of Figure \ref{spectra_single_dual_corona}, with the {\tt vkompthdk} model plotted in blue in the middle and bottom panels (obs no. 6).  As it appears from the Figure, {\tt vkompthdk} still cannot fit the minimum of the phase-lag spectra properly, yielding a chi-square of 415.2 for dof 255 (see Table \ref{table2} for best fit parameters). \\

Next we fit the rms spectra of the QPO with the model {\tt{ vkdualdk*dilution}} \citep{ka20,be22} and the lag spectra of the QPO with the model {\tt vkdualdk}. We tie $kT_{\rm e,1}/kT_{\rm e,2}$ of {\tt vkompthdk} to  $kT_{\rm e}$ of {\tt nthcomp}, $\Gamma_{1}/\Gamma_{2}$ of {\tt vkompthdk} to $\Gamma$ of {\tt nthcomp}, and link all the parameters of  {\tt vkompthdk} between the lag spectra and the rms spectra. We allowed the seed photon temperature of the first corona, $kT_{\rm s,1}$, to vary independently, while tying the seed photon temperature of the second corona, $kT_{\rm s,2}$, to the inner disk temperature, $kT_{in}$. The chi-square for the fit with a dual-corona model decreases to 254.2 for dof 251 (obs no. 6). We give the best-fitting parameters for obs no. 6 in Table \ref{table2} along with the chi-square of the fits for both models. In the middle and bottom panels of Figure \ref{spectra_single_dual_corona}, the red colour solid line represents the best fit model of {\tt vkdualdk*dilution} and {\tt vkdualdk} to, respectively, the rms and phase-lag spectra of the QPO. The corner plot of the spectral parameters for obs no. 6, generated using {\tt pyXspecCorner}\footnote{\url{https://github.com/garciafederico/pyXspecCorner}}, is shown in Figure \ref{corner_plot}. During the period MJD 60181--60199, the phase-lag spectra at the QPO frequency are flat (1.0 Hz < $\nu_{QPO}$ < 2.5 Hz). The phase lag spectra are reasonably well-fitted by the single-corona model {\tt vkompthdk}; however, when a valley is evident in the lag spectra (see Figure \ref{phase_lag_all}), fits using the dual-corona model, {\tt vkdualdk}, provide statistically better results.

For most observation in the period MJD 60199--60203 (obs.~28--42), the dual-corona model appears to over-fit the data, while the single-corona model provides statistically acceptable fits. For instance, in observation 32, the single-corona model yields $\chi^2$/dof = 115.7/153. In these cases, the over-fitting of the dual-corona model may be explained by the decreasing QPO rms amplitude over time or, in some cases, by shorter exposure times, which result in the rms and lag spectra being less able to constrain the model parameters. Nevertheless, even in cases where the dual-corona model over-fit the data, the sizes of the two coronas remain consistent with each other, similar to the earlier period. For the few observations in this period where the dual-corona model still provides a statistically better fit (see Tables \ref{table3}--\ref{table6}), the sizes of the two coronas again remain in good agreement. To avoid inconsistencies in our analysis by switching between single and dual-corona models, we opt to fit all observations in with the dual-corona model. In cases where the dual-corona model over-fits the data, we link the sizes of the two coronas to be the same, ensuring a coherent and systematic approach to the modeling.\\

After the data gap between observations 42 (MJD 60203.4) and 43 (MJD 60219.7), the size of the second corona can no longer be constrained when fitting with the dual-corona model. We initially tied the parameters of the two coronas, but since the lag is not flat in these observations, unlike in previous ones, the single-corona model could not produce the valley in the lag spectrum. First, we tied the sizes of both coronas ($L_1$ and $L_2$), which resulted in a reduced chi-square of 143.0/174 as shown in Figure \ref{app_spectra_single_dual_corona} (blue solid line). While  this is statistically a good fit, the model still did not reproduce the observed phase lag valley. Next, we let the two corona sizes vary separately, but we were unable to constrain $L_2$, likely due to the limited number of bins in those observations. We then fixed $L_2$ to the value obtained in observation 39, the last one where the second corona
can be reliably measured ($L_2 = 1.9 \times 10^3$ km). This gives a $\chi^2/dof$ of 128.5/173 and  the model fits the valley of the phase lag spectra as shown by the red solid line in Figure \ref{app_spectra_single_dual_corona}.

From the fits to the time-averaged spectrum of the source and the rms and phase lag spectra of the QPO, we find that during the hard and hard-intermediate states,  the inner disk temperature, $kT_{\rm in}$, increases from $\sim$0.27 keV to $\sim$0.73 keV with time as shown in the left panel of Figure \ref{qpo_para_radio}. The power-law index increases from $\sim$1.6 to $\sim$3.0, which highlights the evolution of the source from the hard to the hard-intermediate state during the course of the observations, as shown in the right panel of Figure \ref{qpo_para_radio}. The coronal seed temperature, $kT_{\rm s,1}$ also increases from $\sim$0.25 keV to $\sim$0.70 keV. We show the evolution of $kT_{\rm s,1}$ in the top left panel of Figure \ref{qpo_para_radio}. The $kT_{\rm in}$,
$kT_{\rm s,1}$ and power-law index increase and then decrease as a function of time between MJD 60199 and MJD 60201.\\

 Using the parameters of {\tt vkdualdk} we have computed the intrinsic feedback fraction, the fraction of the corona flux, $\eta_{\rm{int,1}}$ and $\eta_{\rm{int,2}}$, that returns to the disk (see \citealt{ka20} for details) from corona 1 ($C_1$) and corona 2 ($C_2$), respectively (given in Appendix Tables \ref{table3}--\ref{table6}). We show $\eta_{\rm{int,1}}$ and  $\eta_{\rm{int,2}}$ as a function of time with the black and blue points in  the bottom left panel of Figure \ref{qpo_para_radio}, respectively. The intrinsic feedback fraction for both coronas is $\sim 2-5$\% when the QPO frequency is in the range 0.3--2.6 Hz. When the QPO frequency is above 2.6 Hz, the feedback fraction rises to $\sim 15-20$\% for corona 1 and to $\sim$10\% for corona 2. Note that we have frozen the value of the feedback fraction in obs. 17, because it remained unconstrained otherwise.

Assuming that the source is at a distance of 2.7 kpc \citep{mata_2024} and has an inclination angle between $30-60^{\circ}$ \citep{ve23}, we translate the {\tt diskbb} norm parameter (given in Appendix Tables \ref{table3}--\ref{table6}) to an inner disk radius, including a spectral hardening factor
$T_{col}/T_{eff}$ of 1.7 \citep{sh95,ku98}. Assuming that the system harbours a 10 M$_\odot$ black hole, we compute the inner disk radii range in units of the gravitational radius, $R_g= GM/c^{2}$, as shown in the bottom right panel of Figure \ref{qpo_para_radio} as a function of time. The shaded green region in Figure  \ref{qpo_para_radio} shows that the inner disk radius decreases from 30--40 $R_g$ to 5--7 $R_g$ as the QPO frequency increases from 0.3 to 7.0 Hz.\\

The analysis reveals that, as the QPO frequency increases from 0.3 Hz to 2.5 Hz, $L_1$ decreases from $\sim 6.5 \times 10^3$ km to $\sim 2 \times 10^3$ km with some fluctuations in between as shown in Figure \ref{corona_size_radio}. During the first observation, $L_2$ starts at approximately $\sim 2 \times 10^3$ km, expands to around 
$\sim 8 \times 10^3$ km, and then contracts to nearly the same values as $L_1$. For the first observation $L_2$ is $ \sim 2 \times 10^3$ km, it then expands to $\sim 8 \times 10^3$ km and then contracts to a size that is nearly equal to $L_1$. This trend is also consistent with the results of \citet{liao_24}, which fit the {\tt vkompthdk} model between MJDs 60198 to 60203 and find that the size of corona decreases from $\sim$2500 km to $\sim$1000 km and then increases again to $\sim$2000 km. After a gap of about two weeks in the data, the QPO is at frequency of $\sim6$ Hz. Assuming no further variation in the size of the second corona in the observations with QPO frequency in the 6--7 Hz range, which we fixed to $1.9 \times 10^3$ km (see previous paragraph), the size of the first corona increases again, from $\sim 2 \times 10^3$ km to $\sim 10 \times 10^3$ km at 6.0 Hz, and further to $\sim 15 \times 10^3$ km when the QPO frequency reaches 7.0 Hz. These findings are summarized in Tables \ref{table3}--\ref{table6}, where values fixed during the fitting process are indicated with square brackets. We provide all the parameters of both coronas for all the observations in Tables \ref{table3}--\ref{table6}.
\section{Discussion}
\label{sec:dis}

We analyse NICER observations of \swiftJ{} during the hard and hard-intermediate states of its first outburst, covering the period from MJD 60181 to 60226 (late August to mid-October 2023). The source shows type-C QPOs with centroid frequencies between 0.3 to 7.0 Hz. We model the time-averaged spectra of the source, as well as the rms and phase-lag spectra of the QPO, using the time-dependent Comptonization model {\tt vkompth}. At the start of the observations, approximately 90\% of the total $2-20$ keV flux comes from the corona and 5\% comes from the disk, while by the end of the observations, the contribution of the corona and disk are 75\% and 20\%, respectively (see rows 6 and 10 of Table \ref{table6}). Our results indicate the presence of two horizontally extended coronas ($C_1$ and $C_2$ from now on) 
covering a truncated accretion disk. As the source progresses towards the HIMS, the inner disk moves closer to the black hole. Initially, $C_1$ and $C_2$ contract; later, $C_1$ expands. The temperature of the seed-photon source for both coronae increases from 0.3 keV to 0.7 keV, suggesting the corona shifts inwards along with the disk. In an alternative scenario, the corona remains around 10 $R_g$ from the black hole while the disk moves inward, such that seed photons that illuminate come progressively from regions of the disk closer to the black hole. We discuss the evolution of the corona and other spectral parameters in the context of radio observations in Section \ref{dis:corona}.\\

The power-law index,  $\Gamma$, reaches a value of $\sim$3 by the end of these observations. While $\Gamma \ge$ 2.5 is typically associated with the soft-intermediate and soft states, the occurrence of Type-C QPOs, the slope of the power spectrum, the high amplitude of the broadband noise component, and the dominance of the non-thermal flux (comprising 80\% of the total flux) indicate that in all these observations the source is in the hard-intermediate state. 
Furthermore, using the Power colour-colour diagram, and following
the methodology outlined by \citet{heil_2015}, \citet{ingram_2024} classified these observations as being in the hard-intermediate state, as shown in Figure 7 of their paper.
 Additionally, a photon index of $\Gamma \sim$ 2.9 has been reported previously for the hard-intermediate state in the black hole binary GRS 1915+105 \citep{Droulans_2009}.\\

\subsection{QPO lag spectrum: Evidence for a Hard-to-Soft lag transition}
At lower QPO frequencies, $\nu_{QPO}$$\sim 0.3$ Hz, the lag spectrum displays a hard lag, where the 2.0--10 keV photons lag behind the 0.5--2.0 keV band photons by approximately 0.3 rad, or 150 ms. For QPO frequencies in the range 0.8 Hz < $\nu_{QPO}\le$ 1.2 Hz, the lag spectrum gradually flattens, with the lags fluctuating at around $-0.1$ rad and +0.1 rad. Here, the time lag between the 2.0–10 keV and 0.5–2.0 keV bands decreases progressively from about 15 ms down to 2 ms. As the QPO frequency exceeds 1.2 Hz, the lag spectrum transitions to a soft lag, where the 0.5–2.0 keV photons lag behind the 2.0–10 keV photons. In this regime, the lag values shift to negative $-1$ ms at 1.2 Hz to $-8$ ms at 2.6 Hz, and further decrease, reaching values between $-10$ ms and $-12$ ms for frequencies in the range 6.2~Hz~<~$\nu_{QPO}$~<~7.0~Hz. A similar transition of the QPO lag spectra from hard to soft as a function of QPO frequency has been previously observed in other black hole X-ray binaries, such as GRS 1915+105, around a QPO frequency of 2 Hz (see Figure 2 of \citealt{Reig_2000} and Figure 4 of \citealt{zh20}) and for MAXI J1535$-$571 at a QPO frequency of 2.2 Hz \citep{garg_2022}.\\

The QPO fractional rms amplitude increases from 7.6\% to 9.3\% as the QPO frequency increase from 0.34 Hz to 0.8 Hz. The sizes of the coronas are $L_1 \approx 6.5 \times 10^3$ km and $L_2 \approx \sim 2 \times 10^3$ km. As the QPO frequency increases further from 0.8 Hz to 2.6 Hz, the QPO fractional rms amplitude decreases from 9.3\% to 6.1\%. Both coronas shrink to $\sim$2 $\times$ $10^3$ km at this point. For a QPO frequency in the range $6-7$ Hz the fractional rms amplitude further decreases from 2\% to 1\%; in this regime the two coronas separate themselves with $C_1$  expands from $\sim$2 $\times 10^3$ km to $\sim$$10^4$ km and we assumed that the $C_2$ remains of the size $\sim$2 $\times 10^3$ km. From the model fits, $\eta_{int}$ for both coronas is initially around 1-2\% and then stays at 5\% for both. A $\eta_{int}$ of 5\% suggests a horizontally extended corona, as a vertically extended corona would result in a much lower $\eta_{int}$ \citep{ga22}. By the end of the observation, $\eta_{int}$ increases to 15–20\% for $C_1$ and 10\% for $C_2$, indicating that while the two coronas remain separated, they are still horizontally extended. In the LHS/HIMS of MAXI J1820+070, \citet{ma23} found that the two coronas gives similar values $\eta_{int}$ \citep[see Figure 9 of][]{ma23} and explained it through horizontally extended coronas. The phase lag between the two coronas, $\phi$, remains consistently close to $\pi$ rad (see $\phi$ in Appendix Table \ref{table3}, \ref{table4}, \ref{table5} and \ref{table6}. Note that $\phi$ is defined cyclically in the range $(-\pi, \pi]$). The fact that the two coronas are in anti-phase suggests that the time variations of the external heating rate $-$explicitly included in {\tt vkompth}$-$ and hence the energy, are transferred from one corona to the other over the course of the QPO cycle. Remarkably, this behaviour persists even as the QPO frequency changes by a factor of $\sim 20$, pointing to a robust mechanism that couples the two regions over a wide range of timescales.
\subsection{Dual corona}
\label{dis:corona}
Our spectral-timing results indicate that on MJD 60181, as the X-ray flux increases (see Figure \ref{lightcurve}), $L_1 \approx 6.5 \times 10^3$ km and $L_2 \approx 2 \times 10^3$ km, while the inner disk is truncated at $\sim$40 $R_g$. The seed-photon temperature for $C_1$ is always lower than $C_2$. The intrinsic feedback fraction of 2\% for both coronas implies that both coronas are horizontally extended and covering (part of) the accertion disk. During the same period, on MJD 60181.98, radio observations with the Very Large Array (VLA) at 5.25 and 7.45 GHz frequencies, shows that the source has an inverted spectral index ($S_\nu$ $\propto$ $\nu^{\alpha}$, with $\alpha$ =0.2), which implies the presence of  a compact jet in the hard state \citep{miller_2023_atel}.\\

By MJD 60184, the disk moves closer to the black hole to $\sim$30 $R_g$, the seed-photon temperatures for $C_1$  and $C_2$ increase to $\sim$ 0.34 keV and 0.30 keV, respectively and both coronas expands to $\sim$7 $\times 10^3$ km. The VLITE observations centred at MJD 60185.99 show a significant radio brightening (see left panel of Figure \ref{lightcurve}), confirmed by the Allen Telescope Array \citep[ATA;][]{bright_2023_atel}. The inner disk radius moves to $\sim$25 $R_g$ and both coronas shrink further to $\sim$3 $\times 10^3$ km. The seed-photon temperature of $C_1$ is always lower than that of $C_2$  (see Table \ref{table3}). From MJD 60186.08-- 60193.94 an extended continuous jet, which became fainter and less extended while the core brightness remained relatively constant, has been reported \citep{wood_2024}. The sudden increase of the seed-photon temperature of $C_1$ could be associated with a discrete southern jet knot \citep{wood_2024}, which occurs at nearly the same time. \citet{wood_2024} propose that this knot may arise from downstream internal shocks or jet-interstellar medium interactions over transient relativistic jets. A comprehensive investigation into the impact of these shocks on the disk-corona system would require detailed radio and X-ray simulations, which is beyond the scope of this study.\\ 

Between MJD 60199 and MJD 60201, the QPO frequency, inner disk temperature, seed-photon temperature, power-law index, and inner disk radius all exhibit abrupt increases followed by declines. This marks the transition of the source to a relatively softer state before it fully transitioned to the high-soft state. At MJD 60202 the inner disk radius decreases to $\sim$10 Rg, while the corona size remains more or less constant at nearly $\sim$2 $\times 10^3$ km. The seed-photon temperature for $C_1$ is $\sim$0.3 keV and the seed-photon temperature for $C_2$ is $\sim$0.45 keV, which marks the separation of both coronas, although the sizes are nearly the same. Using ART-XC observations, \citet{Mereminskiy_24} reported a transition from a type-C into type-B QPO at MJD 60205 just before a giant X-ray flare reported by MAXI at MJD 60206.5 followed by a radio flare peak at MJD 60209.99 (see Figure \ref{lightcurve}). Since NICER did not observe the source during this period, we are unable to determine any changes in coronal properties during this period.\\

Between MJD 60219 and MJD 60222, the inner disk radii remains at $\sim$5 $R_g$, $C_1$ expands again reaching a size of nearly $\sim$$10^4$ km and $C_2$ being at the $\sim$2 $\times 10^3$ km (we fix it as we could not constrain it.). The intrinsic feedback fraction of $C_1$ increases to $\sim$20\% while that of $C_2$ is $\sim$10\%. A second radio flare, observed by the VLA at 5.25 GHz, shows a rapid increase of the compact jet flux density, from 7.2 $\pm$ 0.1 mJy to 235.8 $\pm$ 1.1 mJy  within a day \citep[MJD 60222-60223;][]{miller_2023b_atel}. The quenching of the compact radio jet, and the subsequent dramatic radio flare with flattening of the spectral index, strongly suggest the ejection of transient, relativistic jets \citep{fender_2004}. Through X-ray hardness and intensity with MAXI, \citet{mata_2024} claimed that the source made short transitions to the soft-intermediate state during the same period. If these state transitions are indeed linked to jet ejections, then the Comptonizing medium could serve as the source of the ejected material as proposed by \citet{Rodriguez_2003} for the microquasar source XTE~J1550$-$564 and by \citet{me22} for GRS 1915+105. Due to the lower significance of the QPO in this period, we are unable to assess whether there are changes in the coronal or disk properties associated with the radio flare or the transition to the soft-intermediate state in this work.\\

Later, on 14 October (MJD 60231.540) a bright flare was detected with the  RATAN-600 radio telescope with a flux of 770 $\pm$ 30 mJy at 11.2 GHz \citep{Trush_2023ATel}, which could potentially signify that the source transitioned to the SIMS \citep{fender_2004}. Our results show that before transitioning to the SIMS, the size of corona 1 expands from   $\sim$2 $\times 10^3$ km to $\sim$$10^4$ km. A similar horizontal expansion in the coronal size, staying parallel to the accretion disk, was reported by \citet{ma23} for the BHXB source MAXI J1820+070 in the HIMS using the {\tt vkompth} model \citep[see Figure 11 of ][]{ma23}. It should be noted that the {\tt vkompth} model assumes a spherical coronal geometry; however, the true coronal geometry may be more complex. Therefore, the corona size reported here should be interpreted as a characteristic size rather than the exact physical size \citep[see][]{me22,ga22}.\\

For a horizontally extended corona, the expected polarization angle is aligned with the jet axis \citep{Poutanen_1996,ursini_2022}. Assuming the radio
polarization aligns with the jet direction, \citet{ingram_2024} reported that the radio polarization is consistent with the X-ray polarization angle (PA $\sim$ 2$^\circ$) for \swift{}. The alignment of the PA with the radio jet direction also indicates that the corona is horizontally extended. Our proposed geometry of the corona is consistent with the polarimetry results obtained with IXPE \citep{ingram_2024} for \swift{}. The seed-photon temperature of the two coronas differs, with $C_1$ consistently having a lower seed-photon temperature than $C_2$. This suggests that $C_1$ covers the outer parts of the disk, while $C_2$ covers the inner regions, with some overlap, as their temperatures differ by 100 eV only. Here, $C_1$ primarily influences the rms and lag spectra at the QPO frequency, while $C_2$ dominates the time-averaged spectrum.
The fact that the two coronal temperature were required to reproduce the observed phase lag spectra suggests that the two coronas overlap. 
\label{sec:con}
\begin{acknowledgements}
We thank the referee for their constructive suggestions and comments which has improved the quality of the manuscript. This research has made use of NICER data and HEASOFT software provided by the High Energy Astrophysics Science Archive Research Center (HEASARC), which is a service of the Astrophysics Science Division at NASA/GSFC. DR acknowledges financial support from Centre National d'\'Etudes
 Spatiales (CNES) and the Science Survey Center of XMM-Newton.  MM acknowledges support from the research programme Athena
with project number 184.034.002, which is (partly) financed by the Dutch Research Council (NWO). FG is a CONICET Researcher. FG acknowledges support from PIP 0113 and PIBAA 1275 (CONICET).
\end{acknowledgements}
\bibliographystyle{aa}
\bibliography{aanda}
\appendix
\section{Figures and tables}
\begin{figure*}
    \rotatebox{-90}{\includegraphics[scale=0.6]{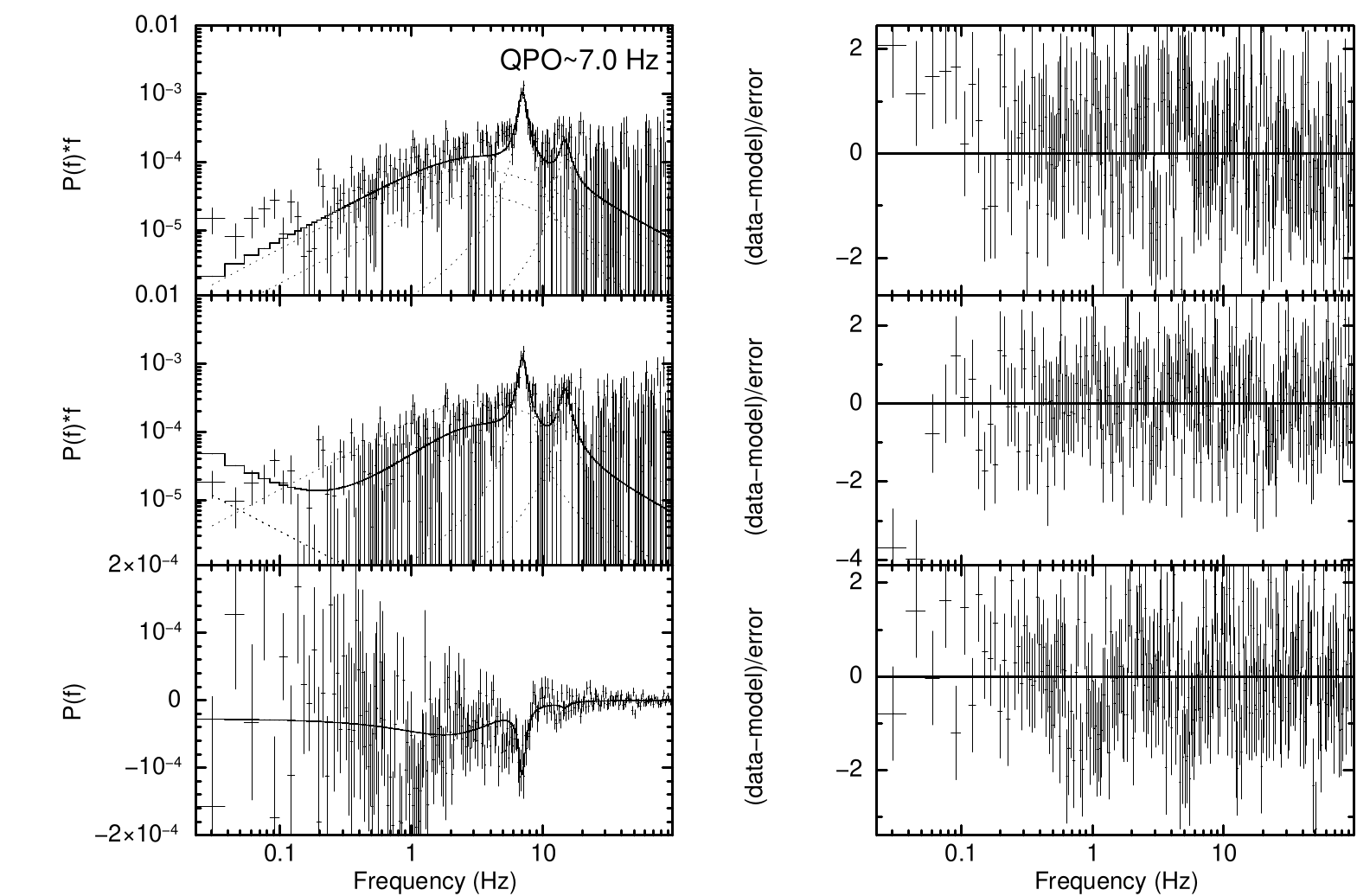}}
\caption{Same as Figure \ref{PDS_real_imag} but for observation ID 6557020402 when the type-C QPO centroid frequency was 7.0 Hz.}
\label{app_PDS_real_imag}
\end{figure*}

\begin{figure}[h!]
    \centering
    \includegraphics[scale=0.5]{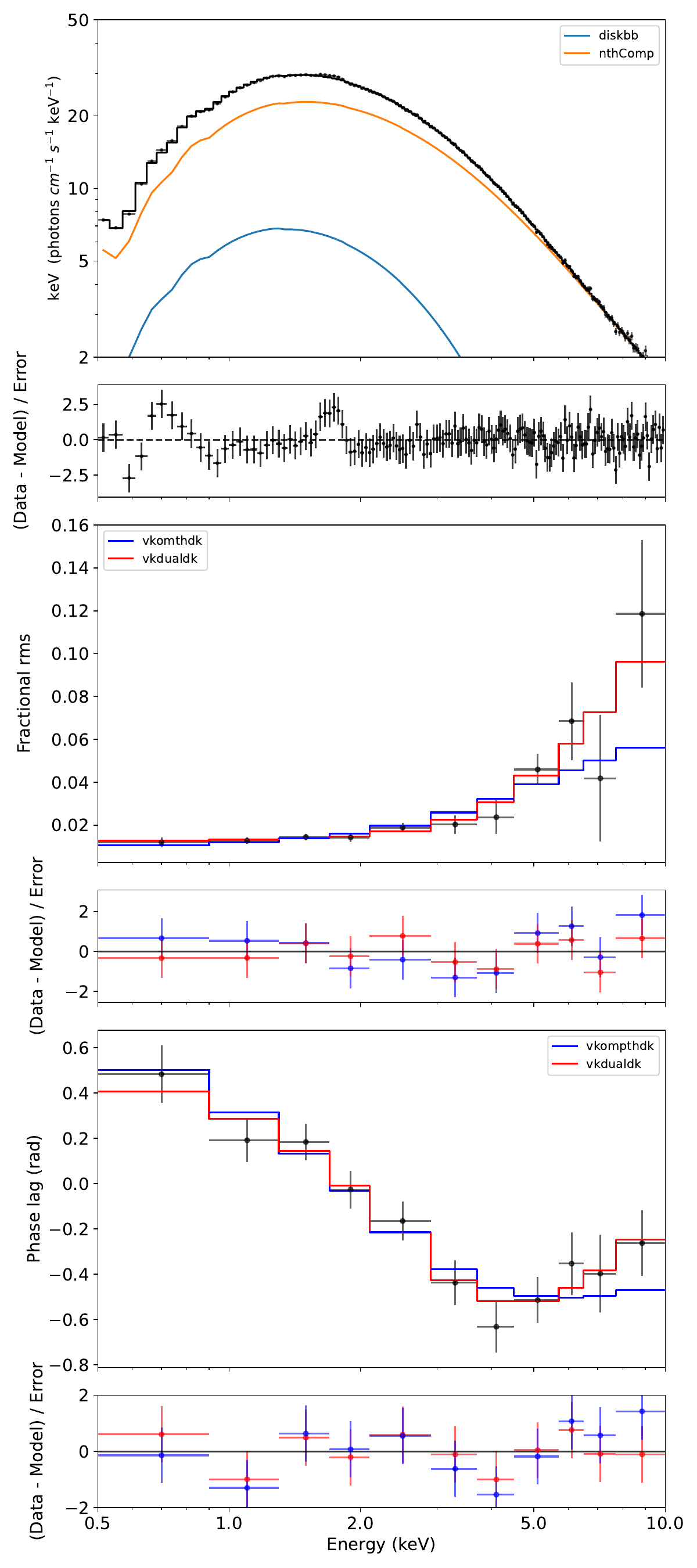}
    \caption{Same as Figure \ref{spectra_single_dual_corona} but for observation ID 6557020402 with the type-C QPO at a centroid frequency 7.0 Hz.}
    \label{app_spectra_single_dual_corona}
\end{figure}

\begin{figure*}
    \centering{\includegraphics[scale=1.2]{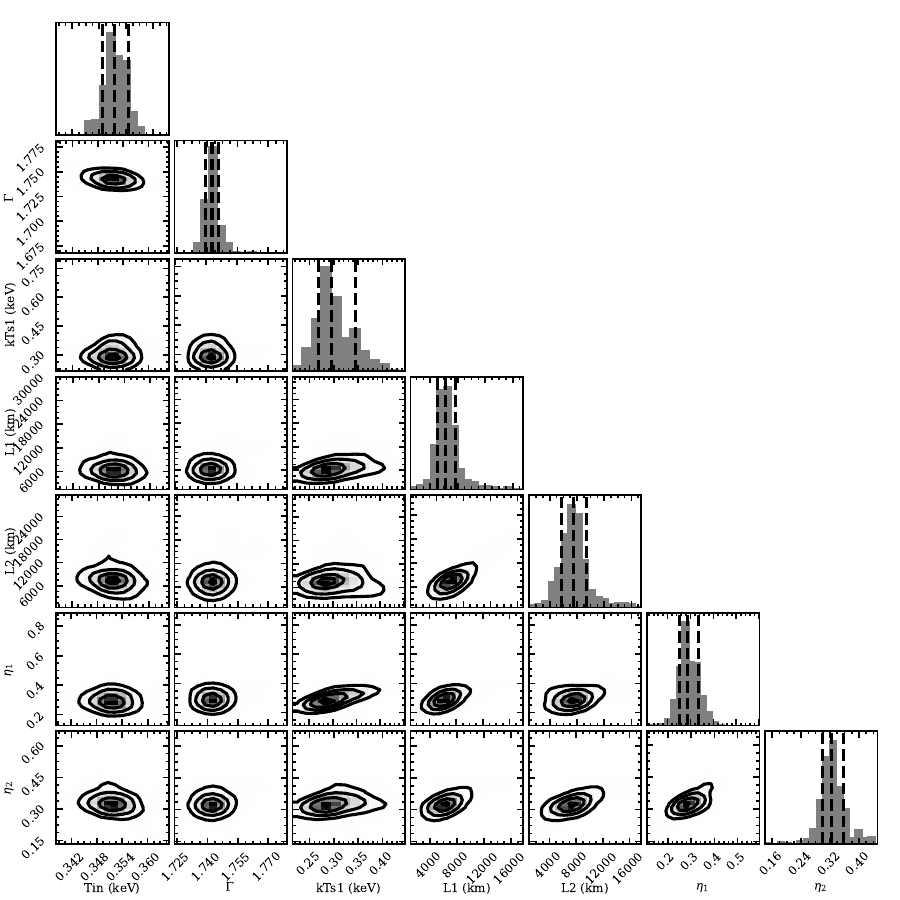}}
\caption{Corner plot for of the parameters of \swiftJ{} for ObsID 6203980105 fitted using  {\tt vkdualdk}. }
\label{corner_plot}
\end{figure*}

We show NICER observations details in Table \ref{table1}, which lists the observation number, NICER OBSID, mid-time MJD, and exposure in kiloseconds. We present the fit to the PDS and the real and imaginary components of the cross spectrum for OBSID 6557020402 in Figure \ref{app_PDS_real_imag}. The panels are similar to the Figure \ref{PDS_real_imag}. Figure \ref{app_spectra_single_dual_corona} displays the time-averaged spectrum of the source fitted with the model {\tt TBfeo*(diskbb + nthComp)} (top panel), along with the rms and phase-lag spectra of the 7.0 Hz QPO, fitted with the models {\tt vkompthdk*dilution} and {\tt vkompthdk}, respectively (middle and bottom panels). The {\tt vkompthdk} model is shown in blue. The red colour solid line in middle and bottom panels represent the best fit model of {\tt vkdualdk*dilution} and {\tt vkdualdk} to, respectively, the rms and phase-lag spectra of the QPO. 
\begin{table}[h]
\caption{NICER observation log of \swiftJ: The columns list the observation number, NICER ObsID, mid-time MJD, exposure in kilo seconds, $t_{\rm exp}$, the 0.5--10.0 keV count rate, the QPO frequency and QPO fractional rms amplitude. }
\begin{center}
\scalebox{0.65}{%
\begin{tabular}{ccccccc}
\hline
Obs  & ObsID  &   Time & $t_{\rm exp}$ &  count rate & QPO  frequency & QPO Fractional \\ 
&  &  (MJD mean) &(ks) & ($10^{2}$ counts s$^{-1}$) &  (Hz) & rms (\%)  \\ 
\hline 
1 &  $^{*}$6203980101  &  60181.38  &  0.5  &  153.3 $\pm$ 0.8  & 0.357 $\pm$ 0.011  & 2.9$^\star$  \\
2 & 6203980102 & 60182.64 & 1.1 & 492.9 $\pm$ 0.1 & 0.334 $\pm$ 0.003 & 7.6 $\pm$ 0.3 \\
3 & 6203980103  &  60183.38  & 0.5  &  550.1 $\pm$ 0.3  & 
 0.454 $\pm$ 0.006  &  8.0 $\pm$ 0.8  \\
4 &  6203980103  &  60183.59  &  0.5  &  618.4 $\pm$ 0.3 &  0.604 $\pm$ 0.004  &  8.8 $\pm$ 0.4 \\
5 & 6203980104 & 60184.52 & 3.1 & 732.1 $\pm$ 0.1 & 0.788 $\pm$ 0.003 & 9.3 $\pm$ 0.2 \\
6 & 6203980105 & 60185.49 & 3.8 & 747.5 $\pm$ 0.1 & 0.887 $\pm$ 0.002 & 8.9 $\pm$ 0.1 \\
7 & 6703010101 &  60185.65 & 3.4 & 741.1 $\pm$ 0.1 & 0.867 $\pm$ 0.003 & 9.2 $\pm$ 0.2 \\
8 & 6203980106 &  60186.46 & 10.6 & 783.9 $\pm$ 0.1 & 1.111 $\pm$ 0.002 & 8.6 $\pm$ 0.1 \\
9 &  6750010101  &  60186.97  &  0.6  &  783.0 $\pm$ 0.2  &  1.187 $\pm$ 0.004  &  8.4 $\pm$ 0.3 \\
10  &  6750010102  &  60187.26  &  7.4  &  775.8 $\pm$ 0.1  &  1.123 $\pm$ 0.002  & 8.3 $\pm$ 0.1  \\
11 & 6203980107 &  60187.75 & 5.4 & 770.8 $\pm$ 0.1 &  1.189 $\pm$ 0.003 & 8.0 $\pm$ 0.2 \\
12 & 6203980108 &   60188.49 & 9.4 & 771.1 $\pm$ 0.1 &  1.239 $\pm$ 0.002 & 8.2 $\pm$ 0.1 \\
13 & 6703010102 &  60188.65 & 3.7 & 767.0 $\pm$ 0.1 &  1.216 $\pm$ 0.002 & 8.5 $\pm$ 0.2 \\
14 & 6203980109 &   60189.49 & 12.6 & 755.7 $\pm$ 0.1 & 1.261 $\pm$ 0.001 & 8.5 $\pm$ 0.1 \\
15  &  6750010201  &  60190.20  &  7.2  &  756.2 $\pm$ 0.1  & 1.307 $\pm$ 0.002  &  8.2 $\pm$ 0.1  \\
16 & 6203980110 &  60190.75 &  6.7 & 746.8 $\pm$ 0.1 & 1.304 $\pm$ 0.002 & 8.0 $\pm$ 0.1 \\
17 & 6203980111 &  60191.49 & 9.8 & 703.1 $\pm$ 0.1 & 1.178 $\pm$ 0.001 & 8.3 $\pm$ 0.1 \\
18 & 6703010103 &  60191.42 & 2.8 & 699.2 $\pm$ 0.1 & 1.127 $\pm$ 0.004 & 8.1 $\pm$ 0.3 \\
19 & 6203980112 &  60192.52 & 6.8 & 707.7 $\pm$ 0.1 & 1.424 $\pm$ 0.002 & 7.3 $\pm$ 0.1 \\
20 &  6750010301  &  60193.46  &  7.3  &  708.0 $\pm$ 0.1  &  1.427 $\pm$ 0.002  &  8.0 $\pm$ 0.1  \\
21 & 6203980113 &  60193.68 & 5.4 & 701.5 $\pm$ 0.1 & 1.398 $\pm$ 0.002 & 7.6 $\pm$ 0.1 \\
22 & 6203980114 &  60194.32 & 7.4 & 693.2 $\pm$ 0.1 & 1.478 $\pm$ 0.010 & 6.2 $\pm$ 0.3 \\
23 & 6703010104 &   60194.71 & 0.8 & 714.0 $\pm$ 0.2 & 1.622 $\pm$ 0.006 & 7.4 $\pm$ 0.2 \\
24 &  6750010501  &  60194.87  &  2.3  &  682.7 $\pm$ 0.1  &  1.448 $\pm$ 0.004  &  8.0 $\pm$ 0.1  \\
25 & 6750010502 & 60195.29 &  4.8 &  661.6 $\pm$ 0.1  & 1.358 $\pm$ 0.002 & 7.9 $\pm$ 0.1 \\
26  &  6203980115  &  60195.77  &  1.3  &  647.4 $\pm$ 0.2  & 1.352 $\pm$ 0.005  &  7.3 $\pm$ 0.2 \\
27  &  6203980116  &  60196.13  &  0.5  &   648.0 $\pm$ 0.2  &  1.397 $\pm$ 0.010  &  7.0 $\pm$ 0.3  \\
28  & 6703010106 & 60199.15 & 0.2 & 685.1 $\pm$ 0.2  & 2.267 $\pm$ 0.010 & 5.9 $\pm$ 0.3 \\
29 & 6703010106 & 60199.28 & 0.4 &  737.5 $\pm$ 0.2  & 2.760 $\pm$ 0.008 & 5.5 $\pm$ 0.2 \\
30 & 6203980118 & 60199.60 & 0.8 &   705.1 $\pm$ 0.2  & 2.455 $\pm$ 0.008 & 5.7 $\pm$ 0.2 \\
31 & 6203980118 & 60199.93 & 1.1 &   775.7 $\pm$ 0.2  & 3.154 $\pm$ 0.007 & 5.4 $\pm$ 0.1 \\
32  & 6511080101   & 60200.06 & 0.6 &  793.2 $\pm$ 0.3  & 3.378 $\pm$ 0.008 & 5.0 $\pm$ 0.1 \\
33 & 6511080101   & 60200.12 &  0.6 &  735.2 $\pm$ 0.3   & 2.827  $\pm$ 0.008 & 5.3 $\pm$ 0.1 \\
34 & 6203980119 & 60200.28 & 2.1&  723.5 $\pm$ 0.1  & 2.704 $\pm$ 0.005 & 5.7 $\pm$ 0.1 \\
35 & 6203980119 & 60200.60 & 1.4 &  688.1 $\pm$ 0.1  & 2.606 $\pm$ 0.004 & 5.7 $\pm$ 0.1 \\
36 & 6203980119 & 60200.90 & 1.3 &  628.7 $\pm$ 0.1  & 2.056 $\pm$ 0.004 & 6.3 $\pm$ 0.1 \\
37 & 6203980120  & 60201.51 & 6.3 &  607.6 $\pm$ 0.1  & 1.992 $\pm$ 0.003 & 6.3 $\pm$ 0.1 \\
38 &  6203980121  &  60202.12  &  1.7  &  636.8 $\pm$ 0.2  &   2.308 $\pm$ 0.004  &  5.7 $\pm$ 0.1  \\
39 & 6703010107 &  60202.63 & 4.5 & 678.9 $\pm$ 0.1 & 2.555 $\pm$ 0.004 & 6.1 $\pm$ 0.1 \\
40 &  6750010202  &  60202.58  &   6.5  &  669.2 $\pm$ 0.1  &  2.740 $\pm$ 0.003  &  5.5 $\pm$ 0.1  \\
41 & 6750010203 & 60203.19 & 1.9 &  641.7 $\pm$ 0.2  & 2.480 $\pm$ 0.004 & 5.7 $\pm$ 0.1 \\
42 &  6203980122  &  60203.43  &  1.0  &  686.6 $\pm$ 0.2  &   2.531 $\pm$ 0.012  &  5.2 $\pm$ 0.2   \\
43 & 6203980130 &  60219.70 & 2.6 & 666.9 $\pm$ 0.1 & 6.195 $\pm$ 0.022 & 2.1 $\pm$ 0.1 \\
44 & 6203980131 &  60220.12 & 0.8 & 680.2 $\pm$ 0.2 & 6.856 $\pm$ 0.031 & 1.7 $\pm$ 0.1 \\
45 & 6703010113 &  60220.64 & 3.5 & 677.2 $\pm$ 0.1 & 6.913 $\pm$ 0.021 & 1.6 $\pm$ 0.1 \\
46 & 6557020401 &  60221.57 & 6.5 & 663.2 $\pm$ 0.1 & 6.840 $\pm$ 0.014 & 1.6 $\pm$ 0.1 \\
47 & 6557020402 &  60222.09 & 0.6 & 652.5 $\pm$ 0.2 & 7.067 $\pm$ 0.032 & 1.5 $\pm$ 0.1 \\
48 &  $^{*}$6703010115  &  60222.54  &  1.3   &  707.8 $\pm$ 0.2  &   7.157 $\pm$ 0.175 &   0.7$^\star$   \\
49 &  $^{*}$6203980132  &  60222.67  &  2.3  &  714.9 $\pm$ 0.1  &   8.207 $\pm$ 0.241 &   0.4$^\star$  \\
50 &  $^{*}$6203980133  &  60223.51  &  5.2  &  697.1 $\pm$ 0.1  &    [8.2]$^{a}$  &    0.2$^\star$  \\
51 &  $^{*}$6203980134  &  60224.51  &  3.9  &  679.0 $\pm$ 0.1  &   8.179 $\pm$ 0.209 &    0.3$^\star$  \\
52 &  $^{*}$6203980135  &  60225.54  &  0.6  &  663.8 $\pm$ 0.2  &     [8.2]$^{a}$  &    0.3$^\star$  \\
53 &  $^{*}$6703010116  &  60225.45  &  3.1  &  666.0 $\pm$ 0.1  &    [8.2]$^{a}$ &    0.1$^\star$  \\
54 & 6203980136 &  60226.35 & 2.0 & 577.6 $\pm$ 0.1 & 6.697 $\pm$ 0.027 & 1.1 $\pm$ 0.1 \\
\hline  
\end{tabular}
}
\label{table1}
\end{center}
\tablefoot{All uncertainties are provided at the 1 $\sigma$ confidence level.
\tablefoottext{*}{Observations with QPO's significance less than 3-$\sigma$.}
\tablefoottext{$\star$}{3-$\sigma$ upper limit of the parameter.}
\tablefoottext{a}{fixed parameter during the fitting.}
}
\end{table}
 Figure \ref{corner_plot} shows corner plot of the parameters for \swift{} in ObsID 6203980105, fitted with the {\tt vkdualdk} model. The contours represent the 1$\sigma$, 2$\sigma$, and 3$\sigma$ confidence levels for pairs of parameters, while the vertical lines indicate the 1$\sigma$ confidence intervals for individual parameters.
Tables \ref{table3}, \ref{table4}, \ref{table5} and \ref{table6} give the time-averaged spectral and corona model parameters of \swift{} for observations 1 to 54. The rows include the observation number, QPO frequency, time lag of the QPO between the 0.5–2.0 keV and 2.0–10.0 keV bands, hydrogen column density ($N_{\rm H}$), power-law photon index of {\sc{nthcomp}} ($\Gamma$), inner disk temperature ($kT_{in}$), seed photon temperature of {\sc{vkdualdk}} ($kT_{s,1}$), sizes of the two coronae ($L_1$ and $L_2$), fraction of flux from the seed photon source due to feedback from the corona ($\eta_1$ and $\eta_2$), as well as the chi-square and degrees of freedom of the fit. Errors are given at 1$\sigma$ confidence levels.
\begin{table*}
        \renewcommand\arraystretch{1.5}  
	\centering
	\caption{Time-averaged spectral and corona model parameters of \swiftJ{}. }
	\label{table3}
	\begin{threeparttable}
	\resizebox{\textwidth}{!}{
	\begin{tabular}{lcccccccccccc} 
	\hline
 & &  & & & & Observation number& &  & & & &\\
	Component & Parameter & 2 & 5 & 6 & 7 & 8 & 9 & 10 & 11 & 12 & 13 &14 \\
	\hline
	& QPO frequency (Hz) & 
     0.334 $\pm$ 0.003  &
0.788 $\pm$ 0.003 &
0.887 $\pm$ 0.002 &
0.867 $\pm$ 0.003 &
1.111 $\pm$ 0.002 &
1.187 $\pm$ 0.004 &
1.123 $\pm$ 0.002 &
1.189 $\pm$ 0.003 &
1.239 $\pm$ 0.002 &
1.216 $\pm$ 0.002 &
1.261 $\pm$ 0.001  
 \\
& time lag ($10^{-2}$ sec) & 
 $ 15.49 \pm 2.0 $ &
$ 2.12 \pm 0.27 $ &
$ 2.12 \pm 0.19 $ &
$ 1.55 \pm 0.18 $ &
$ 0.31 \pm 0.08 $ &
$ -0.14 \pm 0.2 $ &
$ 0.26 \pm 0.1 $ &
$ 0.46 \pm 0.1 $ &
$ -0.03 \pm 0.06 $ &
$ -0.06 \pm 0.08 $ &
$ -0.13 \pm 0.04 $ \\
TBfeo& $N_{\rm H}\,(10^{22}\,\rm{cm^{-2}})$ & 
    [0.28]$^{a}$ &
    [0.28]$^{a}$ &
    [0.28]$^{a}$ &
    [0.28]$^{a}$ & 
    [0.28]$^{a}$ & 
    [0.28]$^{a}$ &
    [0.28]$^{a}$ & 
    [0.28]$^{a}$ & 
    [0.28]$^{a}$ & 
    [0.28]$^{a}$ & 
    [0.28]$^{a}$ \\ 
    diskbb & $kT_{\rm in}$ (keV) &
    {$0.270\pm{0.001}$} &
{$0.339\pm{0.003}$} &
{$0.352\pm{0.003}$} &
{$0.350\pm{0.001}$} &
{$0.376\pm{0.003}$} &
{$0.396\pm{0.004}$} &
{$0.389\pm{0.004}$} &
{$0.386\pm{0.003}$} &
{$0.389\pm{0.003}$} &
{$0.388\pm{0.003}$} &
{$0.393\pm{0.003}$} 
  \\
     & ${N_{\rm disk}}^{b}\,(10^{4})$ &
{$31.7\pm{1.0}$} &
{$17.8\pm{0.4}$} &
{$15.7\pm{0.4}$} &
{$15.9\pm{0.2}$} &
{$12.1\pm{0.2}$} &
{$10.3\pm{0.2}$} &
{$10.6\pm{0.2}$} &
{$10.9\pm{0.2}$} &
{$10.3\pm{0.2}$} &
{$10.5\pm{0.2}$} &
{$9.9\pm{0.2}$}\\ 
     & ${F_{\rm disk}}^{*}\,(\rm{10^{-8}~ergs~cm^{-2}~s^{-1}})$ &
 { 3.4 $\pm$ 0.1 } &
{ 4.8 $\pm$ 0.1 } &
{ 4.8 $\pm$ 0.1 } &
{ 4.8 $\pm$ 0.1 } &
{ 5.0 $\pm$ 0.1 } &
{ 5.2 $\pm$ 0.1 } &
{ 5.0 $\pm$ 0.0 } &
{ 5.0 $\pm$ 0.1 } &
{ 4.8 $\pm$ 0.1 } &
{ 4.9 $\pm$ 0.1 } &
{ 4.9 $\pm$ 0.1 }     \\      
    nthComp & $\Gamma$ & 
{$1.621\pm{0.005}$} &
{$1.721\pm{0.006}$} &
{$1.742\pm{0.003}$} &
{$1.735\pm{0.005}$} &
{$1.799\pm{0.003}$} &
{$1.817\pm{0.011}$} &
{$1.803\pm{0.003}$} &
{$1.812\pm{0.004}$} &
{$1.835\pm{0.004}$} &
{$1.828\pm{0.003}$} &
{$1.832\pm{0.003}$}    
 \\ 
     & $kT_{\rm e}$ (keV) & {[40]$^{a}$} &
{[40]$^{a}$} &
{[40]$^{a}$} &
{[40]$^{a}$} &
{[40]$^{a}$} &
{[40]$^{a}$} &
{[40]$^{a}$} &
{[40]$^{a}$} &
{[40]$^{a}$} &
{[40]$^{a}$} &
{[40]$^{a}$} \\ 
     & $N_{\rm nthComp}$ &
{$23.5\pm{0.1}$} &
{$33.6\pm{0.3}$} &
{$34.4\pm{0.3}$} &
{$33.9\pm{0.2}$} &
{$36.2\pm{0.3}$} &
{$35.5\pm{0.5}$} &
{$35.5\pm{0.3}$} &
{$35.6\pm{0.3}$} &
{$36.0\pm{0.3}$} &
{$35.5\pm{0.3}$} &
{$34.8\pm{0.3}$}
\\
     & ${F_{\rm nthComp}}^{*}\,(\rm{10^{-8}~ergs~cm^{-2}~s^{-1}})$ &
 { 56.7 $\pm$ 0.7 } &
{ 62.6 $\pm$ 0.6 } &
{ 60.9 $\pm$ 0.2 } &
{ 61.3 $\pm$ 0.5 } &
{ 56.7 $\pm$ 0.2 } &
{ 54.3 $\pm$ 0.7 } &
{ 55.8 $\pm$ 0.1 } &
{ 54.5 $\pm$ 0.2 } &
{ 52.4 $\pm$ 0.2 } &
{ 52.5 $\pm$ 0.1 } &
{ 51.3 $\pm$ 0.1 } 
 \\      
    gaussian & $lineE$ (keV) &
    [6.4]$^{a}$ &
    [6.4]$^{a}$ &
    [6.4]$^{a}$ &
    [6.4]$^{a}$ &
    [6.4]$^{a}$ &
    [6.4]$^{a}$ &
    [6.4]$^{a}$ &
    [6.4]$^{a}$ &
    [6.4]$^{a}$ &
    [6.4]$^{a}$ &
    [6.4]$^{a}$  \\ 
     & $\sigma\,{\rm (keV)}$ & 
{$1.3\pm{0.1}$} &
{$1.1\pm{0.1}$} &
{$[1.0]^{a}$} &
{$1.1\pm{0.1}$} &
{$[1.0]^{a}$} &
{$1.0\pm{0.2}$} &
{$[1.0]^{a}$} &
{$[1.0]^{a}$} &
{$1.0\pm{0.1}$} &
{$[1.0]^{a}$} &
{$[1.0]^{a}$}     
 
\\ 
     & Strength $(10^{-1})$  &
{$2.8\pm{0.4}$} &
{$2.9\pm{0.4}$} &
{$[2.0]^{a}$} &
{$2.7\pm{0.4}$} &
{$[2.0]^{a}$} &
{$2.0\pm{0.5}$} &
{$[2.0]^{a}$} &
{$[2.0]^{a}$} &
{$2.0\pm{0.2}$} &
{$[2.0]^{a}$} &
{$[2.0]^{a}$}     

 \\ 
    vkdualdk & $kT_{\rm s,1}$ (keV) & 
{$0.248^{+0.003}_{-0.011}$} &
{$0.299^{+0.032}_{-0.028}$} &
{$0.286^{+0.020}_{-0.055}$} &
{$0.325^{+0.002}_{-0.001}$} &
{$0.301^{+0.021}_{-0.059}$} &
{$0.267^{+0.055}_{-0.069}$} &
{$0.292^{+0.016}_{-0.013}$} &
{$0.301^{+0.128}_{-0.111}$} &
{$0.294^{+0.018}_{-0.030}$} &
{$0.276^{+0.031}_{-0.043}$} &
{$0.240^{+0.014}_{-0.031}$}     
 \\
     & $L_{1}\,{\rm (10^{3}\,km)}$  & 
{$6.5^{+0.2}_{-0.5}$} &
{$7.0^{+0.5}_{-0.4}$} &
{$6.1^{+1.5}_{-1.3}$} &
{$2.7^{+2.2}_{-0.1}$} &
{$5.5^{+0.9}_{-0.7}$} &
{$2.2^{+0.9}_{-1.2}$} &
{$2.7\pm {0.3}$} &
{$3.8\pm {0.3}$} &
{$3.5\pm {0.3}$} &
{$3.5^{+0.5}_{-0.4}$} &
{$2.8\pm {0.3}$} \\ 
     & $L_{2}\,{\rm (10^{3}\,km)}$ & 
{$1.9^{+0.4}_{-0.1}$} &
{$7.9^{+0.5}_{-0.4}$} &
{$7.5^{+2.1}_{-1.7}$} &
{$2.7^{+0.6}_{-1.1}$} &
{$6.9^{+1.2}_{-1.0}$} &
{$3.6\pm {1.1}$} &
{$3.4^{+0.2}_{-0.4}$} &
{$5.3 \pm {0.4}$} &
{$4.4^{+0.5}_{-0.4}$} &
{$4.7^{+0.7}_{-0.6}$} &
{$3.7\pm {0.4}$}     \\ 
     & $\eta_{1}$ &  
{$0.17 \pm 0.01$} &
{$0.27^{+0.01}_{-0.03}$} &
{$0.28^{+0.02}_{-0.04}$} &
{$0.28^{+0.02}_{-0.01}$} &
{$0.28^{+0.01}_{-0.02}$} &
{$0.27^{+0.03}_{-0.10}$} &
{$0.26\pm {0.01}$} &
{$0.26^{+0.10}_{-0.04}$} &
{$0.26^{+0.01}_{-0.02}$} &
{$0.25^{+0.02}_{-0.03}$} &
{$0.23^{+0.01}_{-0.03}$}     
   \\ 
     & $\eta_{2}$ & 
{$0.19 \pm 0.01$} &
{$0.29^{+0.01}_{-0.02}$} &
{$0.32\pm {0.01}$} &
{$0.29^{+0.02}_{-0.03}$} &
{$0.34^{+0.02}_{-0.01}$} &
{$0.39^{+0.05}_{-0.03}$} &
{$0.33\pm {0.01}$} &
{$0.32^{+0.10}_{-0.03}$} &
{$0.33\pm {0.01}$} &
{$0.35^{+0.02}_{-0.01}$} &
{$0.36^{+0.02}_{-0.01}$}     
 \\
& {$\delta \Dot{H}_{\rm ext,1}$} &
{$11.00^{+1.17}_{-1.53}$} &
{$10.80^{+0.95}_{-1.29}$} &
{$8.70^{+1.13}_{-1.35}$} &
{$22.40^{+0.57}_{-2.31}$} &
{$6.90^{+0.79}_{-0.85}$} &
{$9.60^{+3.25}_{-18.87}$} &
{$11.10^{+0.74}_{-1.48}$} &
{$6.00^{+2.09}_{-2.43}$} &
{$8.20^{+0.58}_{-0.71}$} &
{$8.50^{+1.10}_{-2.09}$} &
{$11.20^{+1.32}_{-3.32}$}
\\
& {$\delta \Dot{H}_{\rm ext,2}$} &
{$27.6^{+0.4}_{-11.7}$} &
{$5.6^{+2.2}_{-0.7}$} &
{$2.9^{+0.3}_{-0.7}$} &
{$16.6^{+2.7}_{-2.1}$} &
{$2.0^{+0.7}_{-2.1}$} &
{$1.1^{+0.6}_{-0.5}$} &
{$2.7^{+0.6}_{-0.2}$} &
{$1.4^{+0.6}_{-3.1}$} &
{$2.0^{+0.7}_{-0.4}$} &
{$1.5^{+0.5}_{-0.9}$} &
{$1.1\pm {0.2}$} 

\\

     & $\phi$ ($\rm{rad}$) &
  {$[-3]^a$} &
{$3.10^{+0.01}_{-0.02}$} &
{$3.11\pm {0.01}$} &
{$[-3]^a$} &
{$-3.13^{+0.07}_{-0.01}$} &
{$-3.03^{+0.17}_{-0.07}$} &
{$3.14^{+0.01}_{-0.01}$} &
{$3.12^{+0.04}_{-0.03}$} &
{$-3.12^{+0.03}_{-0.01}$} &
{$-3.08^{+0.06}_{-0.02}$} &
{$-3.05^{+0.07}_{-0.04}$}   
  \\
     & $\rm{reflag}$ $(10^{-2})$ &  
{$29.4\pm {1.3}$} &
{$10.4\pm {0.7}$} &
{$8.2\pm {0.5}$} &
{$2.0\pm {1.9}$} &
{$4.3\pm {0.4}$} &
{$-0.4\pm {1.1}$} &
{$1.3\pm {0.6}$} &
{$5.5\pm {0.9}$} &
{$2.1\pm {0.4}$} &
{$1.5\pm {0.5}$} &
{$1.5\pm {0.3}$}  \\
     & $\eta_{\rm{int,1}}$ & 
 {$0.005 \pm {0.001}$} &
{$0.019\pm{0.004}$} &
{$0.014^{+0.003}_{-0.002}$} &
{$0.014\pm{0.001}$} &
{$0.017^{+0.003}_{-0.002}$} &
{$0.017^{+0.004}_{-0.002}$} &
{$0.016\pm{0.002}$} &
{$0.018^{+0.009}_{-0.003}$} &
{$0.017^{+0.002}_{-0.001}$} & 
{$0.016^{+0.001}_{-0.002}$} &
{$0.014\pm 0.001$} \\
     & $\eta_{\rm{int,2}}$ & 
    {$0.006 \pm {0.001}$} &
{$0.021^{+0.005}_{-0.004}$} &
{$0.017\pm{0.001}$} &
{$0.015\pm{0.001}$} &
{$0.021^{+0.001}_{-0.003}$} &
{$0.026^{+0.002}_{-0.003}$} &
{$0.021\pm{0.001}$} &
{$0.021\pm{0.005}$} &
{$0.023\pm{0.001}$} &
{$0.024\pm 0.001$} &
{$0.025\pm 0.001$} \\
     \hline
    & $\chi^{2}$ (dof) &
{$228.2 \;(241 \;)$} &
{$223.6 \;(248 \;)$} &
{$254.2 \;(251 \;)$} &
{$233.4 \;(242 \;)$} &
{$266.3 \;(251 \;)$} &
{$222.6 \;(229 \;)$} &
{$266.1 \;(251 \;)$} &
{$203.6 \;(242 \;)$} &
{$264.0 \;(249 \;)$} &
{$234.5 \;(248 \;)$} &
{$258.9 \;(254 \;)$}    

 \\ 
\hline
\end{tabular}
}
\tablefoot{The time lag is between the 0.5--2.0 keV and 2.0--10.0 keV band of the QPO. The reference lag, \textit{reflag}, of the model is in the 0.5--10.0 keV energy band. The errors are at 1$\sigma$ confidence levels.
\tablefoottext{a}{Fixed parameter during the fitting.}
\tablefoottext{b}{Parameter tied during fitting.}
\tablefoottext{*}{Unabsorbed flux in the 0.5--10.0\,keV energy band.}
}
\end{threeparttable}
\end{table*}
\begin{table*}
    \renewcommand\arraystretch{1.5}
	\centering
	\caption{Same as Table \ref{table3} for observation numbers 15 to 25.}
  
	\label{table4}
	\begin{threeparttable}
	\resizebox{\textwidth}{!}{
	\begin{tabular}{lcccccccccccc} 
	\hline
 & &  & & & & Observation number& &  & & & &\\
	Component & Parameter  & 15 &16 &17 & 18 & 19 & 20 & 21 & 22 &23 &24 &25  \\
	\hline
	& QPO frequency (Hz) & 1.307 $\pm$ 0.002 &
1.304 $\pm$ 0.002 &
1.178 $\pm$ 0.001 &
1.127 $\pm$ 0.004 &
1.424 $\pm$ 0.002 &
1.427 $\pm$ 0.002 &
1.398 $\pm$ 0.002 &
1.478 $\pm$ 0.01 &
1.622 $\pm$ 0.006 &
1.448 $\pm$ 0.004 &
1.358 $\pm$ 0.002 
    \\
& time lag ($10^{-2}$ sec) &
$ -0.18 \pm 0.05 $ &
$ -0.13 \pm 0.05 $ &
$ 0.19 \pm 0.04 $ &
$ 0.42 \pm 0.13 $ &
$ -0.66 \pm 0.06 $ &
$ -0.41 \pm 0.05 $ &
$ -0.33 \pm 0.06 $ &
$ -0.32 \pm 0.11 $ &
$ -0.67 \pm 0.16 $ &
$ -0.15 \pm 0.09 $ &
$ -0.17 \pm 0.06 $ \\

    TBfeo& $N_{\rm H}\,(10^{22}\,\rm{cm^{-2}})$ & 
    [0.28]$^{a}$ &
    [0.28]$^{a}$ &
    [0.28]$^{a}$ &
    [0.28]$^{a}$ & 
    [0.28]$^{a}$ & 
    [0.28]$^{a}$ &
    [0.28]$^{a}$ & 
    [0.28]$^{a}$ & 
    [0.28]$^{a}$ & 
    [0.28]$^{a}$ & 
    [0.28]$^{a}$ \\ 
    diskbb & $kT_{\rm in}$ (keV) &
    {$0.400\pm{0.003}$} &
{$0.397\pm{0.003}$} &
{$0.381\pm{0.003}$} &
{$0.377\pm{0.003}$} &
{$0.391\pm{0.003}$} &
{$0.402\pm{0.003}$} &
{$0.400\pm{0.002}$} &
{$0.397\pm{0.003}$} &
{$0.418\pm{0.004}$} &
{$0.399\pm{0.003}$} &
{$0.400\pm{0.003}$} 
    \\
 
     & ${N_{\rm disk}}^{b}\,(10^{4})$ & {$9.1\pm{0.2}$} &
{$9.4\pm{0.2}$} &
{$10.9\pm{0.2}$} &
{$11.3\pm{0.2}$} &
{$9.3\pm{0.2}$} &
{$8.6\pm{0.2}$} &
{$8.8\pm{0.1}$} &
{$8.8\pm{0.1}$} &
{$7.1\pm{0.1}$} &
{$8.4\pm{0.2}$} &
{$8.7\pm{0.2}$} 
\\

     & ${F_{\rm disk}}^{*}\,(\rm{10^{-8}~ergs~cm^{-2}~s^{-1}})$ & 
{ 4.8 $\pm$ 0.1 } &
{ 4.8 $\pm$ 0.1 } &
{ 4.7 $\pm$ 0.1 } &
{ 4.7 $\pm$ 0.1 } &
{ 4.5 $\pm$ 0.1 } &
{ 4.6 $\pm$ 0.1 } &
{ 4.6 $\pm$ 0.1 } &
{ 4.5 $\pm$ 0.1 } &
{ 4.5 $\pm$ 0.1 } &
{ 4.4 $\pm$ 0.1 } &
{ 4.6 $\pm$ 0.1 }      \\  
     
    nthComp & $\Gamma$ & 
{$1.857\pm{0.003}$} &
{$1.845\pm{0.003}$} &
{$1.811\pm{0.003}$} &
{$1.803\pm{0.006}$} &
{$1.865\pm{0.003}$} &
{$1.874\pm{0.005}$} &
{$1.868\pm{0.004}$} &
{$1.878\pm{0.003}$} &
{$1.932\pm{0.006}$} &
{$1.894\pm{0.008}$} &
{$1.855\pm{0.005}$} \\
     & $kT_{\rm e}$ (keV) & {[40]$^{a}$} &
{[40]$^{a}$} &
{[40]$^{a}$} &
{[40]$^{a}$} &
{[40]$^{a}$} &
{[40]$^{a}$} &
{[40]$^{a}$} &
{[40]$^{a}$} &
{[40]$^{a}$} &
{[40]$^{a}$} &
{[40]$^{a}$} 
 \\ 
     & $N_{\rm nthComp}$ &
     {$35.1\pm{0.3}$} &
{$34.4\pm{0.3}$} &
{$32.2\pm{0.2}$} &
{$32.0\pm{0.3}$} &
{$33.1\pm{0.3}$} &
{$32.6\pm{0.3}$} &
{$32.3\pm{0.2}$} &
{$32.3\pm{0.3}$} &
{$33.5\pm{0.4}$} &
{$32.0\pm{0.4}$} &
{$29.9\pm{0.3}$} 
   \\
     & ${F_{\rm nthComp}}^{*}\,(\rm{10^{-8}~ergs~cm^{-2}~s^{-1}})$ &
{ 49.1 $\pm$ 0.1 } &
{ 49.4 $\pm$ 0.1 } &
{ 49.2 $\pm$ 0.1 } &
{ 49.6 $\pm$ 0.4 } &
{ 45.1 $\pm$ 0.1 } &
{ 44.2 $\pm$ 0.2 } &
{ 44.2 $\pm$ 0.2 } &
{ 42.9 $\pm$ 0.1 } &
{ 40.8 $\pm$ 0.2 } &
{ 41.5 $\pm$ 0.3 } &
{ 42.0 $\pm$ 0.3 } 
  \\      
    gaussian & $lineE$ (keV) &
    [6.4]$^{a}$ &
    [6.4]$^{a}$ &
    [6.4]$^{a}$ &
    [6.4]$^{a}$ &
    [6.4]$^{a}$ &
    [6.4]$^{a}$ &
    [6.4]$^{a}$ &
    [6.4]$^{a}$ &
    [6.4]$^{a}$ &
    [6.4]$^{a}$ &
    [6.4]$^{a}$  \\ 
     & $\sigma\,{\rm (keV)}$ &   
  {$[1.0]^{a}$} &
{$[1.0]^{a}$} &
{$[1.0]^{a}$} &
{$1.1\pm{0.1}$} &
{$[1.0]^{a}$} &
{$1.0\pm{0.1}$} &
{$1.0\pm{0.1}$} &
{$[1.0]^{a}$} &
{$[1.0]^{a}$} &
{$1.2\pm{0.2}$} &
{$1.0\pm{0.1}$}  \\ 
     & Strength $(10^{-1})$  & 
{$[2.0]^{a}$} &
{$[2.0]^{a}$} &
{$[2.0]^{a}$} &
{$2.2\pm{0.3}$} &
{$[2.0]^{a}$} &
{$1.4\pm{0.2}$} &
{$1.5\pm{0.2}$} &
{$[2.0]^{a}$} &
{$[2.0]^{a}$} &
{$1.9\pm{0.4}$} &
{$1.5\pm{0.2}$} 
   \\ 
    vkdualdk & $kT_{\rm s,1}$ (keV) & 
 {$0.290^{+0.029}_{-0.022}$} &
{$0.306^{+0.018}_{-0.041}$} &
{$0.263^{+0.023}_{-0.013}$} &
{$0.271^{+0.029}_{-0.020}$} &
{$0.262^{+0.012}_{-0.010}$} &
{$0.303^{+0.016}_{-0.028}$} &
{$0.296^{+0.025}_{-0.037}$} &
{$0.319^{+0.052}_{-0.030}$} &
{$0.273^{+0.073}_{-0.035}$} &
{$0.268^{+0.033}_{-0.041}$} &
{$0.281^{+0.033}_{-0.011}$} \\ 
     & $L_{1}\,{\rm (10^{3}\,km)}$  & 
{ $3.5^{+0.5}_{-0.4}$} &
{$4.1^{+0.4}_{-0.7}$} &
{$2.9^{+0.3}_{-0.2}$} &
{$3.4\pm {0.6}$} &
{$2.9\pm {0.3}$} &
{$2.9^{+0.2}_{-0.3}$} &
{$3.5^{+0.3}_{-0.4}$} &
{$3.2^{+0.7}_{-0.5}$} &
{$2.6^{+1.1}_{-0.7}$} &
{$2.5\pm {0.4}$} &
{$2.8^{+0.4}_{-0.3}$} \\ 
     & $L_{2}\,{\rm (10^{3}\,km)}$ &
{$4.8^{+0.7}_{-0.6}$} &
{$5.3^{+0.7}_{-0.6}$} &
{$4.1^{+0.4}_{-0.3}$} &
{$4.5^{+1.1}_{-0.8}$} &
{$3.7\pm {0.4}$} &
{$3.9^{+0.3}_{-0.4}$} &
{$4.5^{+0.5}_{-0.6}$} &
{$4.0^{+0.8}_{-0.7}$} &
{$2.8^{+1.1}_{-0.8}$} &
{$3.4^{+0.5}_{-0.7}$} &
{$3.7^{+0.5}_{-0.2}$} \\ 
     & $\eta_{1}$ &  
{$0.26^{+0.02}_{-0.01}$} &
{$0.27^{+0.01}_{-0.02}$} &
{$0.25^{+0.02}_{-0.01}$} &
{$0.25^{+0.02}_{-0.04}$} &
{$0.25^{+0.02}_{-0.01}$} &
{$0.28^{+0.01}_{-0.02}$} &
{$0.28^{+0.01}_{-0.03}$} &
{$0.28\pm {0.02}$} &
{$0.22^{+0.03}_{-0.05}$} &
{$0.25^{+0.03}_{-0.04}$} &
{$0.26\pm {0.02}$}  \\ 
     & $\eta_{2}$ & 
   {$0.35^{+0.02}_{-0.01}$} &
{$0.34\pm {0.01}$} &
{$0.35\pm {0.01}$} &
{$0.32\pm {0.01}$} &
{$0.36\pm {0.01}$} &
{$0.36\pm {0.01}$} &
{$0.36^{+0.02}_{-0.01}$} &
{$0.33\pm {0.02}$} &
{$0.33^{+0.06}_{-0.03}$} &
{$0.36\pm {0.02}$} &
{$0.34\pm {0.01}$}  \\ 
& {$\delta \Dot{H}_{\rm ext,1}$} &
{$7.40^{+0.82}_{-0.95}$} &
{$6.80^{+0.54}_{-0.73}$} &
{$10.60^{+1.15}_{-0.96}$} &
{$9.30^{+1.35}_{-4.60}$} &
{$6.80^{+0.87}_{-0.88}$} &
{$7.00^{+0.50}_{-0.73}$} &
{$6.70^{+0.64}_{-0.78}$} &
{$5.40^{+0.72}_{-0.90}$} &
{$7.10^{+1.47}_{-2.03}$} &
{$8.10^{+1.28}_{-3.27}$} &
{$9.00^{+0.80}_{-1.48}$}  \\
& {$\delta \Dot{H}_{\rm ext,2}$} &
 {$1.4^{+0.3}_{-0.5}$} &
{$1.7^{+0.7}_{-1.0}$} &
{$1.6^{+0.2}_{-0.7}$} &
{$1.8\pm 0.8$} &
{$1.0\pm {0.1}$} &
{$1.5^{+0.5}_{-0.4}$} &
{$1.4^{+0.5}_{-0.8}$} &
{$1.7^{+0.8}_{-1.6}$} &
{$1.0^{+0.6}_{-0.9}$} &
{$1.1^{+0.3}_{-0.8}$} &
 {$1.6^{+0.3}_{-0.5}$} \\
     & $\phi$ ($\rm{rad}$) &
 {$-3.10^{+0.01}_{-0.03}$} &
{$-3.13^{+0.05}_{-0.01}$} &
{$-3.10\pm {0.02}$} &
{$3.13\pm {0.01}$} &
{$[-3]^a$} &
{$-3.09^{+0.05}_{-0.02}$} &
{$-3.09^{+0.05}_{-0.02}$} &
{$-3.12^{+0.04}_{-0.01}$} &
{$-3.04^{+0.20}_{-0.05}$} &
{$-3.07^{+0.08}_{-0.07}$} &
{$-3.12^{+0.03}_{-0.02}$}  \\
     & $\rm{reflag}$ $(10^{-2})$ &  
{$-0.8\pm 0.4$} &
{$1.4\pm 0.4$} &
{$4.2\pm 0.4$} &
{$5.7\pm 0.7$} &
{$-2.9\pm 0.5$} &
{$-2.0\pm 0.4$} &
{$-0.5\pm 0.5$} &
{$-0.5\pm 0.9$} &
{$-4.1\pm 1.5$} &
{$-1.7\pm 0.8$} &
{$-0.9\pm 0.6$} \\
     & $\eta_{\rm{int,1}}$ & 
{$0.019^{+0.003}_{-0.002}$} &
{$0.019^{+0.003}_{-0.002}$} &
{$0.015\pm 0.001$} &
{$0.014^{+0.003}_{-0.002}$} &
{$0.017\pm 0.001$} &
{$0.021\pm 0.001$} &
{$0.020^{+0.003}_{-0.002}$} &
{$0.021^{+0.004}_{-0.002}$} &
{$0.020^{+0.006}_{-0.003}$} &
{$0.019^{+0.005}_{-0.002}$} &
{$0.018\pm 0.002$}  \\
     & $\eta_{\rm{int,2}}$ & 
 {$0.027\pm 0.001$} &
{$0.025^{+0.001}_{-0.002}$} &
{$0.023\pm 0.001$} &
{$0.020\pm 0.001$} &
{$0.028^{+0.000}_{-0.000}$} &
{$0.029\pm 0.001$} &
{$0.028^{+0.001}_{-0.003}$} &
{$0.026^{+0.002}_{-0.005}$} &
{$0.030^{+0.002}_{-0.003}$} &
{$0.030^{+0.001}_{-0.002}$} &
{$0.026\pm 0.001$} \\
     \hline
    & $\chi^{2}$ (dof) &  
 {$264.2 \;(251 \;)$} &
{$250.0 \;(250 \;)$} &
{$266.5 \;(251 \;)$} &
{$221.6 \;(243 \;)$} &
{$208.2 \;(252 \;)$} &
{$235.2 \;(249 \;)$} &
{$216.4 \;(248 \;)$} &
{$214.9 \;(251 \;)$} &
{$197.2 \;(225 \;)$} &
{$215.3 \;(239 \;)$} &
{$214.9 \;(245 \;)$} \\ 
\hline
\end{tabular}
}
\tablefoot{Rows are defined as in Table~\ref{table3}.}
\end{threeparttable}
\end{table*}
\begin{table*}[h!]
    \renewcommand\arraystretch{1.5}
	\centering
	\caption{Same as Table \ref{table3} for observation numbers 26 to 36.}
	\label{table5}
	\begin{threeparttable}
	\resizebox{\textwidth}{!}{
	\begin{tabular}{lcccccccccccc} 
	\hline
 & &  & & & & Observation number& &  & & & &\\
	Component & Parameter & 26 & 27 & 28 & 29 & 30 & 31 & 32 & 33 & 34 & 35 & 36   \\
	\hline
	& QPO frequency (Hz) & 1.352 $\pm$ 0.005 &
    1.397 $\pm$ 0.01 &
2.267 $\pm$ 0.01 & 
2.76 $\pm$ 0.008 &
2.455 $\pm$ 0.008 &
3.154 $\pm$ 0.007 &
3.378 $\pm$ 0.008 &
2.827 $\pm$ 0.008 &
2.704 $\pm$ 0.005 &
2.606 $\pm$ 0.004 &
2.056 $\pm$ 0.004  \\

& time lag ($10^{-2}$ sec) & 
$ -0.06 \pm 0.14 $ &
$ -0.85 \pm 0.24 $ &
$ -1.1 \pm 0.45 $ &
$ -1.51 \pm 0.32 $ &
$ -0.79 \pm 0.14 $ &
$ -1.18 \pm 0.09 $ &
$ -1.05 \pm 0.16 $ &
$ -0.97 \pm 0.14 $ &
$ -0.77 \pm 0.06 $ &
$ -0.81 \pm 0.06 $ &
$ -0.6 \pm 0.08 $ \\

    TBfeo& $N_{\rm H}\,(10^{22}\,\rm{cm^{-2}})$ & 
    [0.28]$^{a}$ &
    [0.28]$^{a}$ &
    [0.28]$^{a}$ &
    [0.28]$^{a}$ & 
    [0.28]$^{a}$ & 
    [0.28]$^{a}$ &
    [0.28]$^{a}$ & 
    [0.28]$^{a}$ & 
    [0.28]$^{a}$ & 
    [0.28]$^{a}$ & 
    [0.28]$^{a}$ \\ 
    diskbb & $kT_{\rm in}$ (keV) &
    {$0.403\pm{0.004}$} &
{$0.414\pm{0.004}$} &
{$0.490\pm{0.006}$} &
{$0.512\pm{0.006}$} &
{$0.480\pm{0.005}$} &
{$0.530\pm{0.006}$} &
{$0.548\pm{0.008}$} &
{$0.518\pm{0.006}$} &
{$0.502\pm{0.006}$} &
{$0.500\pm{0.006}$} &
{$0.453\pm{0.005}$} 
    
 \\
     & ${N_{\rm disk}}^{b}\,(10^{4})$ & {$8.2\pm{0.2}$} &
{$7.7\pm{0.2}$} &
{$3.8\pm{0.1}$} &
{$2.5\pm{0.1}$} &
{$3.5\pm{0.1}$} &
{$1.9\pm{0.1}$} &
{$1.6\pm{0.2}$} &
{$2.8\pm{0.1}$} &
{$2.7\pm{0.1}$} &
{$2.7\pm{0.1}$} &
{$4.2\pm{0.1}$}\\

     & ${F_{\rm disk}}^{*}\,(\rm{10^{-8}~ergs~cm^{-2}~s^{-1}})$ &  
 { 4.4 $\pm$ 0.1 } &
{ 4.7 $\pm$ 0.1 } &
{ 4.6 $\pm$ 0.3 } &
{ 3.6 $\pm$ 0.3 } &
{ 3.9 $\pm$ 0.2 } &
{ 3.1 $\pm$ 0.3 } &
{ 3.0 $\pm$ 0.5 } &
{ 4.2 $\pm$ 0.3 } &
{ 3.6 $\pm$ 0.2 } &
{ 3.5 $\pm$ 0.2 } &
{ 3.7 $\pm$ 0.2 } \\ 
    nthComp & $\Gamma$ & 
    {$1.870\pm{0.005}$} &
{$1.845\pm{0.009}$} &
{$2.020\pm{0.014}$} &
{$2.148\pm{0.012}$} &
{$2.075\pm{0.010}$} &
{$2.223\pm{0.010}$} &
{$2.256\pm{0.011}$} &
{$2.122\pm{0.013}$} &
{$2.149\pm{0.011}$} &
{$2.172\pm{0.009}$} &
{$2.054\pm{0.007}$} 
    \\
    
     & $kT_{\rm e}$ (keV) & {[40]$^{a}$} &
{[40]$^{a}$} &
{[40]$^{a}$} &
{[40]$^{a}$} &
{[40]$^{a}$} &
{[40]$^{a}$} &
{[40]$^{a}$} &
{[40]$^{a}$} &
{[40]$^{a}$} &
{[40]$^{a}$} &
{[40]$^{a}$} 
 \\ 
     & $N_{\rm nthComp}$ &
     {$29.6\pm{0.4}$} &
{$28.5\pm{0.4}$} &
{$30.2\pm{0.8}$} &
{$35.2\pm{0.8}$} &
{$33.1\pm{0.7}$} &
{$38.8\pm{0.9}$} &
{$40.0\pm{1.4}$} &
{$33.8\pm{0.8}$} &
{$34.8\pm{0.7}$} &
{$33.9\pm{0.7}$} &
{$29.8\pm{0.5}$}
    
   \\
     & ${F_{\rm nthComp}}^{*}\,(\rm{10^{-8}~ergs~cm^{-2}~s^{-1}})$ &
  { 40.3 $\pm$ 0.2 } &
{ 41.6 $\pm$ 0.4 } &
{ 33.8 $\pm$ 0.2 } &
{ 33.2 $\pm$ 0.2 } &
{ 33.5 $\pm$ 0.1 } &
{ 33.9 $\pm$ 0.2 } &
{ 34.3 $\pm$ 0.3 } &
{ 33.3 $\pm$ 0.2 } &
{ 32.4 $\pm$ 0.2 } &
{ 30.6 $\pm$ 0.1 } &
{ 30.3 $\pm$ 0.1 }  \\      
    gaussian & $lineE$ (keV) &
    [6.4]$^{a}$ &
    [6.4]$^{a}$ &
    -- &
    -- &
    -- &
    -- &
    -- &
    -- &
    [6.4]$^{a}$ &
    -- &
    [6.4]$^{a}$  \\ 
     & $\sigma\,{\rm (keV)}$ &  
{$[1.0]^{a}$} &
{$0.7\pm{0.2}$} &
-- &
-- &
-- &
-- &
-- &
-- &
{$0.8\pm{0.3}$} &
-- &
{$[1.0]^{a}$}
  \\ 
     & Strength $(10^{-1})$  &
{$[2.0]^{a}$} &
{$1.0\pm{0.3}$} &
-- &
-- &
-- &
-- &
-- &
-- &
{$0.6\pm{0.3}$} &
-- &
{$[2.0]^{a}$}    
   \\ 
    vkdualdk & $kT_{\rm s,1}$ (keV) & 
 {$0.297^{+0.034}_{-0.042}$} &
{$0.262^{+0.159}_{-0.120}$} &
{$0.284^{+0.052}_{-0.033}$} &
{$0.363^{+0.058}_{-0.044}$} &
{$0.336^{+0.056}_{-0.043}$} &
{$0.482^{+0.026}_{-0.022}$} &
{$0.430^{+0.060}_{-0.074}$} &
{$0.396^{+0.044}_{-0.027}$} &
{$0.382^{+0.021}_{-0.017}$} &
{$0.369^{+0.053}_{-0.106}$} &
{$0.278^{+0.083}_{-0.048}$} \\ 
     & $L_{1}\,{\rm (10^{3}\,km)}$  & 
{ $1.8^{+0.3}_{-0.2}$} &
{$4.1^{+3.9}_{-1.1}$} &
{$1.1^{+0.9}_{-0.5}$} &
{$2.4^{+1.2}_{-0.6}$} &
{$1.9^{+1.4}_{-0.7}$} &
{$2.1^{+0.5}_{-0.4}$} &
{$1.9^{+0.3}_{-0.5}$} &
{$1.1^{+0.5}_{-0.4}$} &
{$1.7\pm {0.3}$} &
{$1.5^{+0.6}_{-0.4}$} &
{$1.5^{+0.5}_{-0.4}$} \\ 
     & $L_{2}\,{\rm (10^{3}\,km)}$ & 
{$2.4^{+0.6}_{-0.4}$} &
{$4.8^{+4.4}_{-2.1}$} &
{$[1.1]^{b}$} &
{$[2.4]^{b}$} &
{$[1.9]^{b}$} &
{$[2.1]^{b}$} &
{$[1.9]^{b}$} &
{$[1.1]^{b}$} &
{$[1.7]^{b}$} &
{$1.4^{+0.7}_{-1.0}$} &
{$1.8\pm {0.5}$}\\ 
     & $\eta_{1}$ &  
 {$0.28^{+0.02}_{-0.04}$} &
{$0.27^{+0.16}_{-0.09}$} &
{$0.22^{+0.07}_{-0.05}$} &
{$0.24^{+0.08}_{-0.10}$} &
{$0.28^{+0.08}_{-0.09}$} &
{$0.34^{+0.08}_{-0.04}$} &
{$0.26^{+0.12}_{-0.02}$} &
{$0.32^{+0.06}_{-0.03}$} &
{$0.32\pm {0.03}$} &
{$0.33^{+0.02}_{-0.06}$} &
{$0.27^{+0.06}_{-0.04}$}  \\ 
     & $\eta_{2}$ & 
{$0.36\pm {0.01}$} &
{$0.38^{+0.20}_{-0.07}$} &
{$0.35^{+0.05}_{-0.18}$} &
{$0.30\pm{0.12}$} &
{$0.36\pm {0.08}$} &
{$0.36^{+0.07}_{-0.12}$} &
{$0.01^{+0.70}_{-0.01}$} &
{$0.41\pm {0.04}$} &
{$0.41\pm {0.03}$} &
{$0.45^{+0.08}_{-0.04}$} &
{$0.44^{+0.03}_{-0.04}$}\\ 
& $\delta \Dot{H}_{\rm ext,1}$ &
{$9.90^{+1.32}_{-2.74}$} &
{$7.80^{+3.77}_{-11.44}$} &
{$5.70^{+2.52}_{-3.77}$} &
{$2.30^{+0.81}_{-0.93}$} &
{$2.40^{+0.84}_{-1.04}$} &
{$2.20^{+0.46}_{-0.57}$} &
{$0.90^{+0.50}_{-1.96}$} &
{$3.50^{+0.70}_{-0.94}$} &
{$3.00^{+0.44}_{-0.46}$} &
{$2.60^{+0.79}_{-1.08}$} &
{$5.80^{+1.50}_{-1.71}$}  \\
& {$\delta \Dot{H}_{\rm ext,2}$} &
{$2.1^{+0.9}_{-0.6}$} &
{$1.0^{+0.6}_{-1.3}$} &
{$0.5^{+0.3}_{-0.2}$} &
{$0.4^{+0.3}_{-0.2}$} &
{$0.4^{+0.3}_{-0.4}$} &
{$1.3^{+0.8}_{-0.6}$} &
{$0.1^{+0.1}_{-0.3}$} &
{$1.1\pm {0.3}$} &
{$0.8\pm {0.2}$} &
{$0.6^{+0.4}_{-1.4}$} &
{$0.6^{+0.2}_{-0.6}$} \\
     & $\phi$ ($\rm{rad}$) &
 {$-3.14\pm {0.01}$} &
{$-3.11^{+0.20}_{-0.11}$} &
{$[-3]^a$} &
{$[-3]^a$} &
{$[-3]^a$} &
{$[-3]^a$} &
{$[-3]^a$} &
{$[-3]^a$} &
{$[-3]^a$} &
{$-2.92^{+0.51}_{-0.16}$} &
{$-2.90^{+0.13}_{-0.16}$} \\
     & $\rm{reflag}$ $(10^{-2})$ &  
 {$-0.8\pm {1.3}$} &
{$-5.6\pm 2.2$} &
{$-9.6 \pm 4.7$} &
{$-12.6 \pm 4.0$} &
{$-10.1 \pm 3.0$} &
{$-16.1 \pm 2.0$} &
{$-13.3\pm 2.9$} &
{$-16.9\pm 3.9$} &
{$-12.6\pm 1.7$} &
{$-12.4\pm 1.5$} &
{$-9.1\pm 1.5$} \\
     & $\eta_{\rm{int,1}}$ & 
{$0.021^{+0.005}_{-0.004}$} &
{$0.018^{+0.006}_{-0.002}$} &
{$0.023^{+0.005}_{-0.004}$} &
{$0.036^{+0.010}_{-0.008}$} &
{$0.035\pm 0.006$} &
{$0.056\pm 0.006$} &
{$0.044^{+0.004}_{-0.002}$} &
{$0.044\pm 0.004$} &
{$0.046\pm 0.003$} &
{$0.050^{+0.011}_{-0.007}$} &
{$0.030\pm 0.004$}  \\
     & $\eta_{\rm{int,2}}$ & 
{$0.028\pm 0.002$} &
{$0.028\pm 0.005$} &
{$0.042^{+0.006}_{-0.007}$} &
{$0.048^{+0.019}_{-0.015}$} &
{$0.048\pm 0.008$} &
{$0.060^{+0.008}_{-0.006}$} &
{$0.002^{+0.047}_{-0.001}$} &
{$0.058\pm 0.005$} &
{$0.061\pm 0.003$} &
{$0.067^{+0.004}_{-0.007}$} &
{$0.054^{+0.002}_{-0.003}$} \\
     \hline
    & $\chi^{2}$ (dof) &  
{$211.1 \;(229 \;)$} &
{$210.3 \;(190 \;)$} &
{$144.1 \;(147 \;)$} &
{$157.9 \;(145 \;)$} &
{$164.7 \;(153 \;)$} &
{$159.9 \;(160 \;)$} &
{$115.7 \;(153 \;)$} &
{$155.7 \;(153 \;)$} &
{$193.5 \;(219 \;)$} &
{$263.4 \;(220 \;)$} &
{$254.4 \;(219 \;)$}
 \\ 
\hline
\end{tabular}
}
\tablefoot{Rows are defined as in Table~\ref{table3}.}
\end{threeparttable}
\end{table*}

\begin{table*}[h!]
     \renewcommand\arraystretch{1.5}
	\centering
 \caption{Same as Table \ref{table3} for observation numbers 37 to 54.}
	\label{table6}
	\begin{threeparttable}
	\resizebox{\textwidth}{!}{
	\begin{tabular}{lccccccccccccc} 
	\hline
  & &  & & & & Observation number& &  & & & &\\
	Component & Parameter & 37& 38 & 39 & 40 & 41 & 42 & 43 & 44 &45 & 46 & 47 & 54\\
	\hline
	& QPO frequency (Hz) & 
    1.992 $\pm$ 0.003 &
2.308 $\pm$ 0.004 &
2.555 $\pm$ 0.004 &
2.74 $\pm$ 0.003 &
2.48 $\pm$ 0.004 &
2.531 $\pm$ 0.012 &
6.195 $\pm$ 0.022 &
6.856 $\pm$ 0.031 &
6.913 $\pm$ 0.021 &
6.84 $\pm$ 0.014 &
7.067 $\pm$ 0.032 &
6.697 $\pm$ 0.027 
 \\
& time lag($10^{-2}$ sec) & 
$ -0.38 \pm 0.06 $ &
$ -0.94 \pm 0.11 $ &
$ -1.01 \pm 0.06 $ &
$ -1.08 \pm 0.05 $ &
$ 0.03 \pm 0.08 $ &
$ -0.51 \pm 0.42 $ &
$ -2.83 \pm 0.17 $ &
$ -3.21 \pm 0.36 $ &
$ -2.91 \pm 0.18 $ &
$ -1.26 \pm 0.06 $ &
$ -1.26 \pm 0.21 $ &
$ -1.05 \pm 0.12 $ \\
    TBfeo& $N_{\rm H}\,(10^{22}\,\rm{cm^{-2}})$ & 
    [0.28]$^{a}$ &
    [0.28]$^{a}$ &
    [0.28]$^{a}$ &
    [0.28]$^{a}$ & 
    [0.28]$^{a}$ & 
    [0.28]$^{a}$ &
    [0.28]$^{a}$ & 
    [0.28]$^{a}$ & 
    [0.28]$^{a}$ & 
    [0.28]$^{a}$ & 
    [0.28]$^{a}$ & 
    [0.28]$^{a}$\\ 
    diskbb & $kT_{\rm in}$ (keV) &
    $0.443\pm{0.003}$ &
$0.463\pm{0.004}$ &
$0.454\pm{0.004}$ &
$0.468\pm{0.004}$ &
$0.454\pm{0.004}$ &
$0.484\pm{0.004}$ &
$0.692\pm{0.005}$ &
$0.751\pm{0.005}$ &
$0.761\pm{0.004}$ &
$0.753\pm{0.004}$ &
$0.736\pm{0.005}$ &
$0.725\pm{0.004}$ \\
     & ${N_{\rm disk}}^{b}\,(10^{4})$ &
$5.0\pm{0.1}$ &
$3.8\pm{0.1}$ &
$3.3\pm{0.1}$ &
$2.9\pm{0.1}$ &
$3.3\pm{0.1}$ &
$3.5\pm{0.1}$ &
$0.5\pm{0.1}$ &
$0.6\pm{0.1}$ &
$0.7\pm{0.1}$ &
$0.7\pm{0.1}$ &
$0.6\pm{0.1}$ &
$0.8\pm{0.1}$ \\ 
     & ${F_{\rm disk}}^{*}\,(\rm{10^{-9}~ergs~cm^{-2}~s^{-1}})$ &
5.4 $\pm$ 0.1 &
4.0 $\pm$ 0.1 &
3.6 $\pm$ 0.1 &
2.9 $\pm$ 0.1 &
2.9 $\pm$ 0.1 &
3.0 $\pm$ 0.2 &
2.4 $\pm$ 0.4 &
4.4 $\pm$ 0.5 &
5.1 $\pm$ 0.3 &
4.7 $\pm$ 0.3 &
3.8 $\pm$ 0.5 &
4.7 $\pm$ 0.3 \\      
    nthComp & $\Gamma$ & $1.994\pm{0.006}$ &
$2.083\pm{0.008}$ &
$2.170\pm{0.006}$ &
$2.183\pm{0.006}$ &
$2.155\pm{0.007}$ &
$2.129\pm{0.009}$ &
$2.966\pm{0.017}$ &
$3.050\pm{0.029}$ &
$3.027\pm{0.020}$ &
$3.056\pm{0.017}$ &
$3.080\pm{0.030}$ &
$3.025\pm{0.023}$ \\ 
     & $kT_{\rm e}$ (keV) & [250]$^{a}$ &
[250]$^{a}$ &
[250]$^{a}$ &
[250]$^{a}$ &
[250]$^{a}$ &
[250]$^{a}$ &
[250]$^{a}$ & 
[250]$^{a}$ &
[250]$^{a}$ &
[250]$^{a}$ &
[250]$^{a}$ &
[250]$^{a}$ \\
     & $N_{\rm nthComp}$ & $28.0\pm{0.3}$ &
$30.7\pm{0.5}$ &
$35.5\pm{0.4}$ &
$34.8\pm{0.5}$ &
$33.2\pm{0.5}$ &
$32.5\pm{0.5}$ &
$34.2\pm{0.9}$ &
$30.7\pm{1.1}$ &
$28.7\pm{0.8}$ &
$28.8\pm{0.7}$ &
$30.5\pm{1.2}$ &
$24.1\pm{0.7}$
          \\
     & ${F_{\rm nthComp}}^{*}\,(\rm{10^{-9}~ergs~cm^{-2}~s^{-1}})$ &   
21.7 $\pm$ 0.1 &
32.2 $\pm$ 0.1 &
31.1 $\pm$ 0.1 &
31.4 $\pm$ 0.1 &
30.7 $\pm$ 0.1 &
30.0 $\pm$ 0.1 &
21.4 $\pm$ 0.3 &
19.9 $\pm$ 0.4 &
18.9 $\pm$ 0.3 &
18.6 $\pm$ 0.3 &
19.1 $\pm$ 0.4 &
15.3 $\pm$ 0.3  \\      
    gaussian & $lineE$ (keV) & [6.4]$^{a}$ &
    [6.4]$^{a}$ &
    [6.4]$^{a}$ &
    [6.4]$^{a}$ &
    [6.4]$^{a}$ &
    -- &
    -- &
    -- &
    -- &
    -- &
    -- &
    -- \\ 

     & $\sigma\,{\rm (keV)}$ & $0.6\pm{0.1}$ &
[0.7]$^{a}$ &
$0.7\pm{0.1}$ &
[0.7]$^{a}$ &
[0.7]$^{a}$ &
-- &
-- &
-- &
-- &
-- &
-- &
--
 \\ 
     & Strength $(10^{-1})$ 
     & $0.6\pm{0.1}$ &
$0.6\pm{0.1}$ &
$0.5\pm{0.1}$ &
$0.5\pm{0.1}$ &
$0.6\pm{0.1}$ &
-- &
-- &
-- &
-- &
-- &
-- &
-- 
\\ 
    vkdualdk & $kT_{\rm s,1}$ (keV) &
    $0.379^{+0.011}_{-0.045}$ &
$0.293^{+0.020}_{-0.018}$ &
$0.341^{+0.038}_{-0.016}$ &
$0.396^{+0.015}_{-0.012}$ &
$0.355^{+0.028}_{-0.030}$ &
$0.267^{+0.072}_{-0.049}$ &
$0.625^{+0.015}_{-0.013}$ &
$0.716^{+0.038}_{-0.028}$ &
$0.713^{+0.017}_{-0.015}$ &
$0.714^{+0.028}_{-0.024}$ &
$0.704^{+0.013}_{-0.026}$ &
$0.691^{+0.037}_{-0.030}$ \\ 
     & $L_{1}\,{\rm (10^{3}\,km)}$  &  
$2.4 \pm 0.4$ &
$0.8 \pm 0.2$ &
$1.8^{+0.5}_{-0.3}$ &
$2.0^{+0.2}_{-0.3}$ &
$2.4^{+0.6}_{-0.5}$ &
$2.7^{+0.9}_{-0.8}$ &
$10.4^{+2.1}_{-1.5}$ &
$9.7^{+6.2}_{-3.5}$ &
$9.2^{+2.8}_{-2.0}$ &
$14.4^{+0.8}_{-4.9}$ &
$14.3^{+1.3}_{-4.9}$ &
$10.8^{+5.1}_{-2.1}$  \\ 
     & $L_{2}\,{\rm (10^{3}\,km)}$ & 
$2.6^{+0.5}_{-0.4}$ &
[0.8]$^{b}$ &
$1.9^{+0.7}_{-0.5}$ &
[2.0]$^{b}$ &
[2.4]$^{b}$ &
[2.7]$^{b}$ &
[1.9]$^{a}$ &
[1.9]$^{a}$ &
[1.9]$^{a}$ &
[1.9]$^{a}$ &
[1.9]$^{a}$ &
[1.9]$^{a}$      
\\ 
     & $\eta_{1}$ &  
$0.25 \pm 0.01 $ &
$0.22 \pm 0.02 $ &
$0.23 \pm 0.02 $ &
$0.31 \pm 0.02 $ &
$0.28 \pm 0.04 $ &
$0.18^{+0.07}_{-0.04}$ &
$0.52^{+0.15}_{-0.10}$ &
$[0.59]^{c} $ &
$0.57^{+0.23}_{-0.14}$ &
$[0.97]^{c} $ &
$[0.93]^{c} $ &
$[0.57]^{c} $ 
      \\
   &  $\eta_{2}$ &  
$0.28 \pm 0.01$ &
$0.35 \pm 0.01$ &
$0.30 \pm 0.01$ &
$0.37^{+0.02}_{-0.01}$ &
$0.36^{+0.03}_{-0.04}$ &
[0.3]$^{a}$ &
[0.3]$^{a}$ &
[0.3]$^{a}$ &
[0.3]$^{a}$ &
[0.3]$^{a}$ &
[0.3]$^{a}$ &
[0.3]$^{a}$ 
\\ 
& {$\delta \Dot{H}_{\rm ext,1}$} &
{$7.60^{+1.59}_{-1.79}$} &
{$6.50^{+1.20}_{-1.60}$} &
{$3.60^{+0.34}_{-0.95}$} &
{$2.70^{+0.23}_{-0.29}$} &
{$2.90^{+0.54}_{-0.48}$} &
{$2.80^{+1.18}_{-2.06}$} &
{$0.30\pm{0.03}$} &
{$0.20^{+0.05}_{-0.06}$} &
{$0.30^{+0.03}_{-0.05}$} &
{$0.30^{+0.05}_{-0.04}$} &
{$0.30\pm{0.04}$} &
{$0.20\pm{0.04}$}  \\
& {$\delta \Dot{H}_{\rm ext,2}$} & {$3.4^{+2.0}_{-1.9}$} &
{$0.9\pm{0.1}$} &
{$0.9^{+0.3}_{-0.5}$} &
{$1.2\pm{0.2}$} &
{$0.9^{+0.4}_{-0.3}$} &
{$0.1^{+0.1}_{-0.3}$} &
{[0.9]$^{a}$} &
{[0.9]$^{a}$} &
{[0.9]$^{a}$} &
{[0.9]$^{a}$} &
{[0.9]$^{a}$} &
{[0.9]$^{a}$} \\
     & $\phi$ ($\rm {rad}$) & 
$3.14 \pm 0.01$ &
[-3]$^{a}$ &
$-3.00^{+0.03}_{-0.11}$ &
[-3]$^{a}$ &
[-3]$^{a}$ &
[-3]$^{a}$ &
$-3.06 \pm 0.06$ &
$-3.09^{+0.14}_{-0.17}$ &
$-2.99^{+0.08}_{-0.09}$ &
$-2.78^{+0.15}_{-0.08}$ &
$-2.78^{+0.14}_{-0.05}$ &
$-2.85^{+0.10}_{-0.17}$
      \\
     & $\rm{reflag}$ $(10^{-2})$ &
$-5.1 \pm 1.1$ &
$-10.4 \pm 2.0$ &
$-11.5 \pm 0.4$ &
$-12.0 \pm 0.9$ &
$-11.7 \pm 1.4$ &
$-9.2 \pm 3.9$ &
$-37.8 \pm 2.2$ &
$-38.2 \pm 3.1$ &
$-39.6 \pm 1.1$ &
$-45.3 \pm 5.5$ &
$-45.6 \pm 5.5$ &
$-41.9 \pm 4.9$ 
 \\
     & $\eta_{\rm{int,1}}$ &
$0.020 \pm 0.002$ &
$0.022 \pm0.001$ &
$0.030^{+0.003}_{-0.002}$ &
$0.040 \pm 0.002$ &
$0.021^{+0.007}_{-0.003}$ &
$0.050^{+0.021}_{-0.017}$ &
$0.157^{+0.032}_{-0.027}$ &
$0.180^{+0.055}_{-0.075}$ &
$0.178^{+0.038}_{-0.048}$ &
$0.192 \pm0.082$ &
$0.200^{+0.068}_{-0.056}$ &
$0.180^{+0.054}_{-0.056}$ \\

     & $\eta_{\rm{int,2}}$ & 
$0.023 \pm0.001$ &
$0.037 \pm0.001$ &
$0.038 \pm 0.002$ &
$0.049 \pm 0.002$ &
$0.044^{+0.003}_{-0.002}$ &
[0.096]$^{a}$ &
[0.089]$^{a}$ &
[0.092]$^{a}$ &
[0.091]$^{a}$ &
[0.092]$^{a}$ &
[0.093]$^{a}$ &
[0.091]$^{a}$ 
 \\
     \hline
    & $\chi^{2}$ (dof) & $267.5 \;(226)$ &
$171.9 \;(166)$ &
$239.2 \;(244)$ &
$210.4 \;(234)$ &
$172.1 \;(222)$ &
$148.6 \;(172)$ &
$140.6 \;(183)$ &
$137.0 \;(174)$ &
$149.5 \;(192)$ &
$119.7 \;(194)$ &
$128.5 \;(173)$ &
$176.4 \;(183)$ 
\\ 
\hline
\end{tabular}
}
\tablefoot{Rows are defined as in Table~\ref{table3}.}
\end{threeparttable}
\end{table*}
\end{document}